\documentclass[acmsmall]{acmart}
\AtBeginDocument{%
  }

\setcopyright{cc}
\setcctype{by}
\acmJournal{PACMPL}
\acmYear{2026} \acmVolume{10} \acmNumber{OOPSLA1} \acmArticle{121}
\acmMonth{4} \acmDOI{10.1145/3798229}

\usepackage{algpseudocodex}
\usepackage{calc}
\usepackage{url}
\usepackage{xspace}
\usepackage[frozencache,cachedir=.]{minted}
  \setminted{fontsize=\footnotesize,linenos}
\usepackage{stackrel}
\usepackage{stmaryrd}
\usepackage{subcaption}
\usepackage{tikz}
\usepackage{mathpartir}
  
  \newcommand{\Rule}[1]{\hyperlink{#1}{\TirName {#1}}}

\newcommand\minic{Mini-C\xspace}

\NewDocumentCommand\fstar{}{F$^\star$\xspace}
\NewDocumentCommand\haclstar{}{HACL$^\star$\xspace}
\NewDocumentCommand\lowstar{}{Low$^\star$\xspace}
\NewDocumentCommand{\li}{v}{\mintinline{rust}{#1}}

\newcommand{\tomini}[5]{\ensuremath{#1 \vdash #4\,
\expandafter\ifx\expandafter\relax\detokenize{#2}\relax
\else\triangleleft\,#2\fi
\leadsto #5}}
\newcommand{\toministd}[3]{\tomini{\Gamma}{#1}{t_{ret}}{#2}{#3}}
\newcommand{\tret}{\ensuremath{t_{ret}}}
\newcommand{\ite}[3]{\ensuremath{\mathsf{if}\; (#1)\;\mathsf{then}\;#2\;\mathsf{else}\;#3}}
\newcommand{\while}[2]{\ensuremath{\mathsf{while}\; (#1)\;#2}}
\newcommand{\mutrel}[2]{\stackrel[#2]{#1}\Rightarrow}

\newcommand{\sref}[1]{Section~\ref{sec:#1}}
\newcommand{\fref}[1]{Figure~\ref{fig:#1}}

\newcommand{\comments}{false}
\usepackage{ifthen}
\ifthenelse{\equal{\comments}{true}}
{
  \DeclareRobustCommand{\af}[1]{ {\begingroup\color{green!60!black}{(Aymeric) #1}\endgroup} }
  \DeclareRobustCommand{\jonathan}[1]{ {\begingroup\color{blue!60!black}{(Jonathan) #1}\endgroup}}
}
{
  \DeclareRobustCommand{\af}[1]{ {} }
  \DeclareRobustCommand{\jonathan}[1]{ {} }
}

\begin{document}

\title{Scylla: Translating an Applicative Subset of C to Safe Rust}

\author{Aymeric Fromherz}
\email{aymeric.fromherz@inria.fr}
\affiliation{%
  \institution{Inria}
  \country{France}
}
\orcid{0000-0003-2642-543X}

\author{Jonathan Protzenko}
\email{protz@google.com}
\affiliation{%
  \institution{Google}
  \country{USA}
}
\orcid{0000-0001-7347-3050}

\begin{abstract}
The popularity of the Rust language continues to explode; yet, many critical codebases remain
authored in C.
Automatically translating C to Rust is
thus an appealing course of action. Several works have gone down this path, handling an
ever-increasing subset of C through a variety of Rust features, such as unsafe. While the prospect
of automation is appealing, producing code that relies on unsafe negates the memory safety
guarantees offered by Rust, and therefore the main advantages of porting existing codebases to
memory-safe languages.
We instead advocate for a different approach, where the programmer iterates on the original C,
gradually making the code more structured until it becomes eligible for compilation to safe Rust.
This means that redesigns and rewrites can be evaluated incrementally for performance and
correctness against existing test suites and production environments.

Compiling structured C to safe Rust relies on the following contributions: a
type-directed translation from (a subset of) C to safe Rust; a novel static analysis based on ``split
trees'' which allows expressing C’s pointer arithmetic using Rust’s slices and splitting operations;
an analysis that infers which borrows need to be mutable; and a compilation strategy for C
pointer types that is compatible with Rust’s distinction between non-owned and owned allocations.
We evaluate our approach on real-world cryptographic libraries,
binary parsers and serializers, and a file compression library. We show that these can be rewritten
to Rust with small refactors of the original C code, and that the resulting Rust
code exhibits similar performance characteristics as the original C code.
As part of our translation process, we also identify and report undefined behaviors in the bzip2
compression library and in Microsoft’s implementation of the FrodoKEM cryptographic primitive.
\end{abstract}



\maketitle

\section{Introduction}

Despite decades of research, memory safety issues
remain prevalent in industrial applications; recent studies by
Google~\cite{chromiummemorysafety} and Microsoft~\cite{microsoftmemorysafety}
estimate that 70\% of security vulnerabilities are related to incorrect
memory handling. To tackle this issue, both companies and governments
now advocate for the use of memory-safe languages for safety-critical
systems, notably, Rust~\cite{whitehouse2024,nsa2022safelang,rustinwindows}.

Combining the high performance and low-level idioms commonly provided by
languages like C and C++ with memory safety by design, the Rust programming
language continues to enjoy record levels of popularity in industry, being
ranked as the ``most admired programming language'' for the eighth year in a
row~\cite{github2023rust,stackoverflow2023rust}, leading several widely used
projects to plan a transition to Rust~\cite{rustlinux,mozilla24rustgui,microsoft2022rust,uutils2025rust,signalrust}.
However, while Rust offers clear benefits for new, clean-slate code, the value
proposition is less clear for existing code. Indeed, it can be hard, especially
in an industrial setting, to justify rewriting code that has been
battle-tested, thoroughly debugged, and exhibits no glaring issues. In those
situations, Rust is perceived as a ``nice to have'' and particularly useful
to develop new features, but certainly not a
business priority~\cite{google24rewrite}.

To help with the transition, several recent projects therefore propose
to \emph{automatically} translate C code to Rust~\cite{c2rust,citrus,corrode}.
To support the entirety of the C language, and thus to be directly applicable
to existing, diverse codebases, these tools however target \emph{unsafe} Rust,
which allows the use of C-like idioms such as unchecked pointers or casts between
representations (``transmutations'', in Rust lingo) at the cost of statically guaranteed
memory safety, therefore defeating the purpose of using safe languages.
The intended workflow is to use the output of the translation as a starting point,
before iteratively rewriting generated Rust code into a safer version; several recent
approaches propose static analyses to assist with this process by, e.g., identifying raw pointers that can be
transformed into regular safe Rust \emph{borrows}~\cite{emre2021translating}, or
reconstructing native Rust abstractions out of low-level representations~\cite{hong2024tag,hong2023concrat}.

Following this workflow however raises several challenges.
First, tooling for refactoring translated Rust code is lacking, and previously proposed analyses
and refactoring passes are scattered across a variety of standalone
projects~\cite{emre2023aliasing,hong2023concrat,hong2023improving,hong2024don,hong2025type}.
Case in point, even c2rust, one of the leading tools for C to Rust translation,
dropped support for its refactoring tool (\texttt{c2rust refactor})
in 2022\footnote{\url{https://github.com/immunant/c2rust/releases/tag/v0.16.0}}.
Second, refactoring generated code is complex and error-prone, oftentimes leading to deep structural
rewrites whose correctness may be hard to test in the \emph{new} (Rust) environment. Indeed,
battle-tested, industrial C projects are commonly equipped with an extensive testing infrastructure,
including performance, compliance, and integration tests, which is a crucial part of software
development and refactoring processes. Modifying familiar C code is easier for developers, and a
much less risky proposition than whole-codebase refactoring of autogenerated unsafe Rust.

In this paper, we advocate for a new approach, where the programmer
incrementally rewrites their C code to give it more structure, until such time as
it becomes eligible to compilation to Rust.
Aiming to make our translation predictable and to generate Rust code close to the
original C sources, we focus on an applicative subset of C: we target code that manipulates
and processes data, performs pointer-arithmetic, has structured control-flow,
and is portable.
In particular, we do not support codebases that rely on gotos, or that leverage object
representation through integer-to-pointer casts, pointer tricks, bitfields,
or untagged unions.

We start from Clang's frontend AST, and translate it to a first language dubbed
``\minic''.
Doing so, we resolve typical C subtleties, such as integer
promotions, integer conversions, assignments-as-expressions, post- vs. pre-increments, and so on.
Then, we define a type-directed translation from \minic to safe Rust, relying on a
novel, type-directed approach to compile structured pointer arithmetic to Rust slices and slice splitting,
followed by several novel post-translation analyses to soundly infer mutability quantifiers, derive traits, and other Rust-specific
features.

We implement our approach in a tool, Scylla (pronounced ``C-like''), which
relies on
Clang
to consume existing C code and output safe Rust; we provide an
overview of Scylla in \fref{scylla-overview}. To fit the \minic subset and
guide the tool, we propose the use of targeted rewritings and annotations on the source C program,
to, e.g., rewrite aliasing patterns incompatible with Rust's borrow checker, or inform Scylla that a
tagged union should be translated to a high-level abstract data type (ADT).

We finally evaluate our approach on several existing C projects,
showcasing that our applicative C subset covers several
security-sensitive applications.
Our case studies include parts of Windows's SymCrypt~\cite{symcrypt}
and of the \haclstar~\cite{haclxn} cryptographic libraries, the core part of the bzip2 compression algorithm~\cite{bzip2},
binary parsers and serializers from the EverParse library~\cite{ramananandro2019everparse}
for the CBOR data format~\cite{cbor-rfc}, and a recent
Microsoft implementation of the FrodoKEM post-quantum cryptographic
primitive~\cite{frodokem}. As part of the
translation process, we also identify and report several undefined behaviors present in the source C
code of both bzip2 and FrodoKEM.

\begin{figure*}
  \centering
  \begin{tikzpicture}[node distance = 4.5cm]
  \node (c) [draw, inner sep = 2mm] { C Code };
  \node[draw, right of=c, xshift=-1.8cm, inner sep = 2mm] (cast) { C AST };
  \node (minic) [draw, right of=cast, inner sep = 2mm] { Mini-C AST };
  \node (rustast) [draw, below of=minic, yshift=2.9cm, inner sep = 2mm] { Rust AST };
  \node (rust) [draw, left of=rustast, xshift=-2.5cm, inner sep = 2mm] { Rust Code };

  \draw[-latex,line width=2pt] (c) -- (cast)
    node[midway, above] { \footnotesize \texttt{clang} };
  \draw[-latex,line width=2pt] (cast) -- (minic)
    node[midway, above] { \footnotesize \S\ref{sec:minic} Explicit Casts, }
    node[midway, below] { \footnotesize ADTs, Tuples, \ldots };
  \draw[-latex,line width=2pt] (minic) -- (rustast)
    node[midway, right, yshift=.15cm] { \footnotesize  \S\ref{sec:split} Eliminating }
    node[midway, right, yshift=-.15cm] { \footnotesize Pointer Arithmetic };
  \draw[-latex,line width=2pt] (rustast) -- (rust)
    node[midway, above] { \footnotesize  \S\ref{sec:mutability} Mutability, }
    node[midway, below] { \footnotesize \S\ref{sec:traits} Trait Inference };
  \end{tikzpicture}
  \caption{Overview of the Scylla Translation}
  \label{fig:scylla-overview}
  \vspace{-4mm}
\end{figure*}

\section{Formalizing \minic}
\label{sec:minic}

We start this paper by formally presenting \minic, the language that we compile to safe Rust in \sref{translate}. We
also describe how to obtain \minic from actual C (as parsed by Clang).

\subsection{The \minic Language}

\begin{figure}[h]
  \centering
  \small
  \[
  \begin{array}[t]{rll}
    t ::= & \mathsf{uint8\_t}, \mathsf{bool}, \ldots & \text{integers, booleans\ldots} \\
         & \mathsf{unit} & \text{empty type} \\
         & \mathsf{size\_t} & \text{pointer size} \\
         & t* & \text{pointer type} \\
         & t (\vec t) & \text{function type} \\
         & \mathsf{struct}\ s & \text{struct type}\\
         & (\vec t) & \text{tuple type} \\
         & \mathsf{variant}\ v & \text{variant type}
         \\\\
   d ::= & \mathsf{struct}\ s \left\{ t_i\,f_i; \ldots t_j\,f_j[n_j]; \right\} & \text{struct definition} \\
         & \mathsf{variant}\ v\ C_i: \vec t_i & \text{variant definition}
         \\\\
   e ::= & \mathsf{return}\;e,\
          \mathsf{if}\; (e)\;\mathsf{then}\;e\;\mathsf{else}\;e,\\
          &
          \mathsf{while}\; (e)\;e,\ \mathsf{continue},\ \mathsf{break} & \text{control-flow} \\
          & x = e & \text{assignment} \\
          & \mathsf{let}\,t\ x (= e)?\,\mathsf{in}\,e & \text{variable declaration}\\
          & \mathsf{let}\,t\ x[n] (= \{ \vec e \})? \,\mathsf{in}\,e & \text{array declaration}\\
        & \mathsf{match}\ e\ \mathsf{with}\ C_i\ \vec v_i \rightarrow e_i & \text{pattern-matching}\\
   & x & \text{variable} \\
     & e\,\bowtie\,e\ \text{with}\ \bowtie\ = +, -, \times, \ldots & \text{binary operators}\\
   & e[e] & \text{array indexing} \\
   & e.f & \text{field selection} \\
   & e.i & \text{tuple field selection} \\
   & f(\vec e) & \text{function call} \\
        & *e & \text{dereference} \\
        & \mathsf{malloc}(t, e) & \text{typed heap allocation} \\
        & \&e & \text{address-taking} \\
        & C\,\vec e & \text{variant value} \\
   & (t)e & \text{cast} \\
   & (e_1, \dots, e_n) & \text{tuple ($n>1$)}\\
   & () & \text{unit value}\\
   & n: t & \text{typed integer constant}\\
  \end{array}
\]
\caption{Grammar of \minic types, statements and expressions. We use $e_1; e_2$ as syntactic sugar
for $\mathsf{let}\,\_=e_1\,\mathsf{in}\,e_2$. Declarations are optionally initialized.}
  \label{fig:c-types}
\end{figure}

\minic (\fref{c-types}) shares many features with C: it contains standard
control-flow constructs such as branching and loops, and
heavily relies on pointers and pointer operations, such
as dereferences or address-taking.
However, unlike C, \minic has a ``no-surprises'' semantics. All integers have a fixed width, and
C concepts such as integer promotion or integer conversions are represented explicity using
casts. Untyped pointers (\li+void *+) are also not permitted. Another key distinction is that \minic
is an expression language, unlike C which distinguishes statements and expressions.
A concrete
consequence is that assignments do not return a value in \minic~-- C constructs such as
\li+e1 = e2 = e3+, or \li!p[i++]! are all desugared in \minic.
Finally, although we omit its typing
rules, \minic has proper handling of the \li+bool+ type, meaning loops and conditionals expect test
expressions at type \li+bool+, instead of \li+int+ like in C.

Beyond these basic constructs, \minic also features high-level constructs such as tuples or ADTs
(abstract data types, also known as variant types), along with pattern-matching expressions to
destruct them. These can be synthesized via a mechanism of custom annotations which indicate the
\emph{intent} of the original C program -- we detail this in \sref{reconstruction}.

We remark that
making \minic an expression language incurs no loss of generality: a
statement-and-expression language can simply be seen as an expression language,
where statements have type \li+unit+, and where the \li+;+ operator has type
\li+unit -> 'a -> 'a+. It does have the advantage, however, of simplifying the
formalism since there is only one syntactic category.

\subsection{A Type-Directed Translation to \minic}

We now aim to generate a \minic representation out of a C program. To do so, we first query the
\texttt{clang} compiler to extract the typed abstract syntax tree (AST) of a given compilation
unit, which we then translate to a \minic typed AST. For conciseness, we omit a formal presentation
of actual C syntax, which we expect is familiar to most readers.

This translation to \minic is challenging for several reasons.
First, as mentioned earlier, \minic is stricter about typing: loop and conditional test expressions
must be at type \li+bool+, and array indices must be at type \li+size_t+.
Second, we need to implement the rules of integer
promotion and integer conversion defined by the C standard in order to materialize all implicit
casts.
Finally, owing to
historical function prototypes in the C standard library, information provided by Clang is sometimes
partial. For instance, in \li{uint32_t* x = malloc (4 * sizeof(uint32_t))}, the right-hand side has
type \li+void*+ according to the C standard, instead of the expected \li{uint32_t*} pointer type.

To address these issues, we introduce a type-directed translation from C to \minic, where each C
statement or expression translates to a typed \minic expression.
Our judgments are of the form
$\Gamma\vdash \mathbf e \leadsto e: t$
to model the
translation of C expression $\mathbf e$ (in bold) to \minic expression $e$ (in
regular font) of type $t$. The judgment is overloaded to deal with
both C expressions and lists of statements, and produces \minic expressions in
both cases. Since types are isomorphic and no meaningful translation takes
place, there is no $\mathbf t$ (in bold) and we simply write $t$ (in regular
font) for both C and \minic.
$\Gamma$ is a type environment: it maps variable (and function) names to their
corresponding types. We assume a special variable
$\mathsf{ret}$ is in $\Gamma$ to account for the return type of the current function.

We sometimes need to perform type conversion to a prescribed type $t$, which we write
$\Gamma \vdash \mathbf e\,\triangleleft\,t \leadsto e$.
This
serves two purposes: one, to implement C's implicit promotion rules (which depend on the type
expected by the context); two, to adjust types when Mini-C enforces a stronger typing
discipline than C (e.g., loop conditionals). We explain promotion and conversion
in \sref{promotion}.

When the expected type $t$ is $\mathsf{unit}$, we omit the $: \mathsf{unit}$ part altogether for a
more concise judgement.
In
practice, our judgment also contains a flag that indicates whether we are in a \emph{value-producing
context}, to compile, e.g., \li!i++!, to either $i = i + 1$ (in statement position) or
$\mathsf{let}\,i_0 = i\,\mathsf{in}\,i = i + 1; i_0$ (in array index position) -- for simplicity, we
omit these verbose, but straightforward rules.

We present selected rules for the translation in Figure~\ref{fig:ctominic}.
Because C permits type aliases via \li+typedef+, a preliminary phase in our implementation replaces
\li+typedef+s with their definitions before applying the rules from \fref{ctominic}. For brevity, we
elide this phase, along with the translation of C types to \minic types, which is straightforward.
Suffices to say that, again, our translation is aware of C's subtleties, such as how
\li+void f(int x[4])+ really means \li+void f(int *x)+.

We now review a few salient rules.
\Rule{VarDecl} introduces a new variable in the context, then translates the
remaining C
statements into an expression that forms the continuation of a let-binding; \Rule{ArrayDecl} is
identical, except it knows about C's array decay, meaning references to $x$ have a C pointer type,
not an array type. \Rule{Ret} leverages the special binding in the environment that holds the return
type of the current function. \Rule{If} and \Rule{While} enforce that conditions
be booleans.
\Rule{Malloc} requires that the original C code be structured as to make the
type of the elements apparent -- failure to do so triggers a warning in modern compilers
anyhow. The resulting \minic $\mathsf{malloc}$ construct is structured and contains a type, along
with a number of \emph{elements} (not bytes). In practice, our implementation supports more than
just this particular form of malloc; we omit extra rules.

\begin{figure*}
  \centering
  \small
  \begin{mathpar}
    \inferrule[Ret]{
      \toministd{\Gamma(\mathsf{ret})}{\mathbf e}{e}
    }{
      \toministd{}{\mathsf{return}\ \mathbf e}{\mathsf{return}\ e}
    }

    \inferrule[VarDecl]{
      \toministd{t}{\mathbf e}{e} \\
      \tomini{\Gamma,x:t}{}{}{\vec{\mathbf s}}{e_2}
    }{
      \tomini{\Gamma}{}{\tret}{\mathbf t\ x = \mathbf e; \vec{\mathbf s}}{\mathsf{let}\ t\,x = e\,\mathsf{in}\,e_2}
    }

    \inferrule[ArrayDecl]{
      \toministd{t}{\mathbf e_i}{e_i} \\
      \tomini{\Gamma,x:t\ast}{}{}{\vec{\mathbf s}}{e_2}
    }{
      \tomini{\Gamma}{}{\tret}{\mathbf t\ x[n] = \left\{ \vec{\mathbf e} \right\}; \vec{\mathbf s}}{\mathsf{let}\ t\,x[n] = \left\{ \vec e \right\}\,\mathsf{in}\,e_2}
    }

    \inferrule[Seq]{
      \tomini{\Gamma}{}{\tret}{\mathbf s}{e_1} \\
      \tomini{\Gamma'}{}{\tret}{\vec{\mathbf s}}{e_2}
    }{\tomini{\Gamma}{}{\tret}{\mathbf s;\ \vec{\mathbf s}}{e_1;\ e_2}}

    \inferrule[If]{
      \tomini{\Gamma}{\mathsf{bool}}{\tret}{\mathbf e}{e} \\
      \tomini{\Gamma}{}{\tret}{\vec {\mathbf s_1}}{e_1} \\
      \tomini{\Gamma}{}{\tret}{\vec{\mathbf s_2}}{e_2}
    }{\tomini{\Gamma}{}{\tret}{\ite{\mathbf e}{\vec {\mathbf s_1}}{\vec {\mathbf s_2}}}{\ite{e}{e_1}{e_2}}}

    \inferrule[While]{
      \tomini{\Gamma}{\mathsf{bool}}{\tret}{\mathbf e}{e_1} \\
      \tomini{\Gamma}{}{\tret}{\vec{\mathbf s}}{e_2}
    }{\tomini{\Gamma}{}{\tret}{\while{\mathbf e}{\vec{\mathbf s}}}{\while{e_1}{e_2}}}

    \inferrule[Assign]{
      \toministd{\Gamma(x)}{\mathbf e}{e}
    }{\tomini{\Gamma}{}{\tret}{x = \mathbf e}{x = e}}

    \inferrule[AddrOf]{\Gamma \vdash \mathbf e \leadsto e: t}{\Gamma \vdash \&\mathbf e \leadsto e: t\ast}

    \inferrule[Deref]{\Gamma \vdash \mathbf e \leadsto e: t\ast}{\Gamma \vdash \ast\mathbf e \leadsto \ast e: t}

    \inferrule[Malloc]{\toministd{\mathsf{size\_t}}{\mathbf e}{e}}{
      \Gamma\vdash \mathsf{malloc}(\mathbf e \times \mathsf{sizeof}(\mathbf t)) \leadsto \mathsf{malloc}(t, e): t\ast
    }

    \inferrule[Index]{
      \Gamma \vdash \mathbf e_1 \leadsto e_1: t\ast \\
      \toministd{\mathsf{size\_t}}{\mathbf e_2}{e_2}
    }{
      \mathbf e_1[\mathbf e_2] \leadsto e_1[e_2]: t
    }

    \inferrule[PointerAdd]{
      \Gamma \vdash \mathbf e_1 \leadsto e_1: t\ast \\
      \toministd{\mathsf{size\_t}}{\mathbf e_2}{e_2}
    }{
      \Gamma \vdash \mathbf e_1 + \mathbf e_2 \leadsto e_1 + e_2: t\ast
    }

    \inferrule[Var]{}{
      \Gamma \vdash x \leadsto x: \Gamma(x)
    }

    \inferrule[Call]{
      \Gamma(f) = t_\mathsf{ret} (t_1, \ldots, t_n) \\
      \toministd{t_i}{\mathbf e_i}{e_i}
    }{
      \Gamma \vdash
        f(\mathbf e_1, \ldots, \mathbf e_n) \leadsto f(e_1, \ldots, e_n)
    }

    \inferrule[Constant]{}{
      \Gamma \vdash n: t \leadsto n: t
    }

    \inferrule[Conversion]{
      \Gamma \vdash \mathbf e \leadsto e: t_1
    }{
      \Gamma \vdash \mathbf e \triangleleft t_2 \leadsto \mathsf{convert}(e, t_1, t_2)
    }

    \inferrule[ConvertEq]{}{\mathsf{convert}(e, t, t) = e}

    \inferrule[ConvertNeq]{t\neq t'}{\mathsf{convert}(e, t, t') = (t')e}

    \inferrule[BinOp]{
      \Gamma \vdash \mathbf e_1 \leadsto e_1: t_1 \\
      \Gamma \vdash \mathbf e_2 \leadsto e_2: t_2 \\
      (t'_1, t'_2, t_\mathsf{ret}) = \mathsf{promote}(\bowtie, t_1, t_2)
    }{
      \Gamma \vdash
      \mathbf e_1 \bowtie \mathbf e_2 \leadsto
          \mathsf{convert}(e_1, t_1, t'_1) \bowtie
          \mathsf{convert}(e_2, t_2, t'_2): t_\mathsf{ret}
    }
  \end{mathpar}
  \caption{Translation from C to \minic, Selected Rules}
  \label{fig:ctominic}
\end{figure*}

\subsection{Implicit Conversions and Integer Promotion}
\label{sec:promotion}
Conversions and integer promotions complicate the semantics of C programs. They are one of
the motivating factors behind the introduction of \minic, before actual translation to Rust.
Conversions change the type of an expression according to the context: for instance, in the
right-hand side of an assignment (type determined by the left-hand side), or as an argument to a
function call (type determined by the function prototype)~\cite[6.3]{c11}.

In our judgments,
$\triangleleft$ materializes conversions that are otherwise implicit in the C standard
(\Rule{Conversion}); if the synthesized (bottom-up) type differs
from the expected (top-down) type, a cast is inserted (\Rule{ConvertNeq}), otherwise, the expression
remains as-is (\Rule{ConvertEq}). Further conversions occur as part of the \emph{usual arithmetic
conversions}~\cite[6.3.1.8]{c11}. We follow the pseudo-algorithm outlined in the C standard, and our
function $\mathsf{promote}$ returns the types that the operands should be promoted to, along with
the corresponding return type.
Because we work with fixed-width integers, we can slightly simplify
the algorithm, as we do not need to deal with types like \li+int+ and
\li+long+ which may have the same size but not the same rank.

We conclude with a brief remark: our translation assumes that the C code is portable and does not
rely on C's \emph{data model} (i.e., choice of widths for particular types). In other words, we
expect the code to have the same behavior regardless of whether \li+long+ is 4 or 8 bytes. Our
implementation detects at configure-time the data model for the architecture the code is running on,
then uses this information to go from, e.g., \li+unsigned int+ to \li+uint32_t+.

\subsection{Synthesizing High-Level Constructs}
\label{sec:reconstruction}

Compared to C, \minic also provides higher-level program constructs, such as ADTs or tuples.
While several previous works investigated the automated detection and reconstruction of such
patterns~\cite{hong2024don,hong2024tag}, we instead rely on user-provided
annotations on type definitions to
identify structs that must be translated to, e.g., ADTs.
We believe this yields a more predictable translation, along with a better developer
experience, since we can error out when the usage of a tagged union does not follow the expected
patterns, rather than silently fail to detect it.
Leveraging these annotations, we
now describe how we perform the needed semantic reconstructions to obtain valid \minic.

\paragraph{{Tagged unions and ADTs}}

While the C language does not provide support for abstract data types, a common pattern is
to emulate them through the use of \emph{tagged unions}. A tagged union is a struct type
that combines a union type, representing the payload of the structure, with a tag, commonly
represented as an integer, which tracks the current state of the union. Concretely,
we consider tagged unions of the shape \li{{ int tag; union { t0 case0; ...; tn caseN }}},
and assume that the tag ranges from 0 to N, and that tag values match the order of the union
cases. We show an example of our handling of tagged unions in
\fref{taggedunion}.
For brevity, we omit the corresponding formal rules; in practice, we
carry a supplemental set of flow-sensitive equations
(to keep track of tag values at a given program point), and a supplemental set
of tag mappings (to keep track of the correspondence between tag values and C
union cases).

When annotated with the appropriate attribute (line 5, left), our tool translates this type definition
to a variant type with constructors \li{Case0 t0, ..., CaseN tn} (line 1-3, right). Creating a tagged union value is
straightforward: when translating a value \li{{.tag = i, .casej = e }}, we first, as a sanity-check,
ensure that the \li{.casej} field corresponds to the $i$-th constructor. If so, we translate
expression $e$ with type $t_j$ to \minic expression $e'$, and generate the value $C_j\, e'$.

Translating accesses to a tagged union requires a bit more bookkeeping. To safely access field \li{casei}
of a tagged union variable $x$, it is necessary to know that the tag of $x$ is currently equal to $i$.
This is commonly checked through a conditional branching of the shape \li{if (x.tag == i) { ... x.casei }}, or a switch,
as is shown on line 15 of our example.
We therefore recognize these code patterns, and translate them to standard
pattern-matching \li{match x with | Ci v -> ...}. We store and propagate the state of tagged union $x$, and
translate all occurrences of \li{x.casei} to the constructor payload $v$ inside the branch (line 7, right);
accesses to different union cases are considered invalid and the translation
errors out.

\begin{figure*}
  \begin{minipage}{0.48\linewidth-2em}
    \begin{minted}[xleftmargin=0.5cm, resetmargins]{c}
  #define CaseU8 0
  #define CaseU16 1

  typedef struct
  __attribute__((annotate("scylla_adt")))
  cases_s {
    uint8_t tag;
    union {
      uint8_t case_uint8;
      uint16_t case_uint16;
    };
  } cases;

  void f(cases s) {
    if (s.tag == CaseU8) {
      uint8_t x = s.case_uint8;
    }
  }
    \end{minted}
  \end{minipage}
  \hspace{2em}\vrule\hspace{2em}%
  \begin{minipage}{0.48\linewidth-2em}%
    \begin{minted}[resetmargins]{rust}
 variant cases
   | case_uint8 (v: uint8_t)
   | case_uint16 (v: uint16_t)

 unit f(cases s) {
   match s with
   | case_uint8 v -> let uint8_t x = v in ()
   | _ -> ()
 }
    \end{minted}
  \end{minipage}
  \caption{Example of tagged union translation from C (left) to \minic (right)}
  \label{fig:taggedunion}
\end{figure*}

\begin{figure*}
  \begin{minipage}{0.48\linewidth-2em}
    \begin{minted}[linenos=false,resetmargins]{c}
    cases s = {
      .tag = CaseU8, { .case_uint16 = 0 }
    };
    // Payload does not match the tag
  \end{minted}
  \end{minipage}
  \hspace{2em}\vrule\hspace{2em}%
  \begin{minipage}{0.48\linewidth-2em}%
    \begin{minted}[linenos=false,resetmargins]{c}
    if (s.tag == CaseU8) {
      uint16_t x = s.case_uint16;
    } // Accessed payload does not match
    // the current union state
  \end{minted}
  \end{minipage}
  \caption{Examples of invalid usages of tagged unions, that will not be translated to \minic}
\end{figure*}

\paragraph{{Tuples}}
Similarly to ADTs, we also offer attributes to translate annotated types to tuples.
A struct type with $n$ fields $t_i\,x_i$ translates
to an n-ary tuple $(t_1, \ldots, t_n)$; as part of the translation, we keep track of which types
are tuple types, and translate corresponding field accesses to tuple field accesses at the
appropriate index. This is useful for both code quality, and accepting a greater subset of programs:
because tuples are typed structurally, they benefit from mut-polymorphism, while structs do not --
we say more in \S\ref{sec:mutability}.

\section{Generating Safe Rust Code}
\label{sec:translate}

\minic provides us with an explicit, entirely type-annotated representation of C programs.
We now show how to translate \minic programs to safe Rust. The chief difficulties are: compiling
away C's pointer arithmetic (Section~\ref{sec:split}), making mutability and aliasing explicit
(Section~\ref{sec:mutability}), and automatically providing idiomatic Rust constructs such
as \emph{traits} (Section~\ref{sec:traits}).

\subsection{Base Translation and Pointer Representation}

We begin with the translation of \minic types to Rust types. The difficulty lies in translating
pointer types, because Rust differentiates between \li+Box<T>+ (owned pointer to a heap-allocated T)
and \li+&T+ (un-owned pointer, \emph{a.k.a.} borrow). Furthermore,
unlike C, Rust distinguishes between a pointer to a single element (e.g. \li+&T+) and a pointer
to multiple elements (e.g. \li+&[T]+). Finally, arrays in Rust are values, and do not decay to
pointers automatically -- they must be explicitly converted.

We adopt the following strategy. By default, every pointer type in C, be it to the stack or to the
heap, to a single element or to multiple elements, compiles to a slice borrow \li+&[T]+ (we infer
mutability automatically in \sref{mutability} and therefore ignore the \li+mut+ qualifier for now).
Concretely, \li{uint8_t x; uint8_t *y = &x;} generates \li{let y: &[u8] ...} in Rust, and \emph{not}
\li{&u8}.

Additionally, we provide heuristics as well as manual annotations to guide
the translation towards \li+Box<T>+ instead of \li+&[T]+ when judicious. Our tool performs a limited
analysis, and can conclude that a function \li+T *create()+ that does not contain references to
globals must be allocating a fresh \li+T+, and therefore translates it to \li+fn create() -> Box<T>+.
This analysis recurses (using a fixed-point computation) within struct and variant definitions, to
determine whether a pointer field becomes a Box, or a borrow -- the latter makes the struct
parameterized over a lifetime.

A consequence of this design choice is that our translation must be type-directed, in order to
insert coercions between arrays and slice borrows (since the former do not decay automatically in
Rust); and to reconcile between those two and instances of \li+Box<T>+ introduced by either the user
or our heuristics. We omit the full syntax of Rust, for brevity, and instead focus on the
salient parts of our translation in \fref{expressions}.

Our rules have the form
$\Gamma \vdash e \triangleleft T \rightsquigarrow E \dashv \Gamma'$, meaning that in Rust
typing environment $\Gamma$, the translation of \minic expression $e$ with expected \emph{Rust} type
$T$ yields Rust expression $E$ and an updated Rust typing environment $\Gamma'$.
The output environment serves to invalidate variables that should no longer be
used (we explain below).
We use $t
\hookrightarrow T$ to translate types from \minic to Rust.

Coercions appear in \Rule{E-Array-Slice}, which
introduces a borrow, thus turning the array into a slice borrow;
and in \Rule{E-Box-Slice}, which operates identically, but for boxed slices. Rules \Rule{E-Slice-Box} and
\Rule{E-Array-Box} are ``reverse conversions'', which convert stack and unknown allocations into
heap ones.
Finally, \Rule{E-Var} models the simple case where a
variable does have the expected type.
In practice, some of the syntax introduced by the conversion rules is unnecessary, as the Rust
compiler is able to add auto-borrows and auto-derefs in many places. Our implementation is aware of
this, and features a nano-pass at the very end that removes superfluous $\&$s and $*$s.

The coercions introduced by conversion rules can however lead to subtle semantic differences,
specifically, those introduced for the ``reverse conversions'' \Rule{E-Slice-Box}
and \Rule{E-Array-Box}.
Consider for instance the following (simplified) Rust program, where
\li{y} must be translated as a
\li{Box<[T]>}, meaning the array $x$ is coerced to a boxed slice.
In this program, the assertion
does not hold, as the call to \li{Box::new} creates a copy of \li{x}.
In a C program however (left), both \li{x} and \li{y} would be the same pointer,
therefore the change to \li{y} on line 3 would also apply to \li{x}. (Generally,
  constructors that take another data structure, or explicit
conversions performed via \li{.into()} in Rust generate a copy.)
  \vspace{1ex}%

\noindent
  \begin{minipage}{0.5\columnwidth-4em}%
    \begin{minted}[resetmargins,linenos=false]{c}
uint8_t x[1] = {0};
uint8_t *y = x;
*y = 1;
assert(*x == 1);
    \end{minted}
  \end{minipage}%
  \hspace{2em}\vrule\hspace{2em}%
  \begin{minipage}{0.5\columnwidth-2em}%
    \begin{minted}[resetmargins,linenos=false]{rust}
let x: [u8; 1] = [0; 1];
let mut y: Box<[u8]> = Box::new(x);
y[0] = 1;
assert!(x[0] == 1)
    \end{minted}
  \end{minipage}%
  \vspace{1ex}

These copies cannot be avoided, as in our translation scheme, it might truly be the case that what
initially started as a stack allocation needs to be promoted to a heap allocation. Furthermore,
Rust provides no way of ``opting out'' of the \li{Copy} trait for base types like arrays of
integers, meaning that Rust will silently perform a copy of $x$ into $y$, while allowing further
modifications to $x$. We therefore leverage our output environment $\Gamma'$ and strip the original variables
from $\Gamma$ when performing the conversion, thus forbidding any further usage of $x$.
If the original C program further relies on \li{x}, our translation will error out, and will ask the
programmer to fix their source code. This is another area where we adopt a ``semi-active''
approach, and declare that some patterns are poor enough, even for C, that they ought to be
touched up before the translation takes place.

\newcommand{\typrelbox}{\stackrel[\oblong]{}\hookrightarrow}
\newcommand{\typrellife}{\stackrel['a]{}\hookrightarrow}
\newcommand{\boxt}{\ensuremath{\mathsf{Box}\langle[T]\rangle}}

\begin{figure*}[h]
  \centering
  \small
  \begin{mathpar}
    \inferrule[E-Index]{
      \Gamma \vdash e_1 \triangleleft \&[T] \rightsquigarrow E_1 \dashv \Gamma' \\
      \Gamma' \vdash e_2 \triangleleft \mathsf{usize} \rightsquigarrow E_2 \dashv \Gamma'' \\
    }{
      \Gamma \vdash e_1[e_2] \triangleleft T \rightsquigarrow E_1[E_2]  \dashv \Gamma''
    }

    \inferrule[E-Deref]{
      \Gamma \vdash e \triangleleft \&[T] \rightsquigarrow E \dashv \Gamma'
    }{
      \Gamma \vdash *e \triangleleft T \rightsquigarrow E[0]  \dashv \Gamma'
    }

    \inferrule[E-Var]{
      x: T \in \Gamma
    }{
      \Gamma \vdash x \triangleleft T \rightsquigarrow x \dashv \Gamma
    }

    \inferrule[E-Stack]{
      t \hookrightarrow T \\
      \Gamma_i \vdash e_i \triangleleft T \rightsquigarrow E_i \dashv \Gamma_{i+1} \\\\
      \Gamma_N, x: [T; N] \vdash e' \triangleleft T' \rightsquigarrow E'\dashv \Gamma
    }{
        \begin{array}{rll}
        \Gamma_0 \vdash &
        \mathsf{let}\,t~x[N] = \{ \vec e \}\,\mathsf{in}\, e' \triangleleft \; T' \cr
        \rightsquigarrow &
        \mathsf{let}\; x: [T; N] = [ \vec E ];\; E' \dashv \Gamma \setminus \left\{ x \right\}
        \end{array}
    }

    \inferrule[E-Heap]{
      t \hookrightarrow T \\\\
      \Gamma, x: \boxt \vdash e' \triangleleft T' \rightsquigarrow E' \dashv \Gamma'
    }{
        \begin{array}{rll}
          \Gamma \vdash & \mathsf{let}\,t *\!x = \mathsf{malloc}(t, N)\,\mathsf{in}\, e' \triangleleft \; T' \cr
      \rightsquigarrow &
      \mathsf{let}\; x: \mathsf{Box}\langle[T]\rangle =
      \mathsf{vec!}[ 0; N ].\mathsf{into\_boxed\_slice}();\; E' \dashv \Gamma' \setminus \left\{ x \right\}
      \end{array}
    }

    \inferrule[E-Box-Slice]{
      x: \mathsf{Box}\langle[T]\rangle \in \Gamma
    }{
      \Gamma \vdash x \triangleleft \&[T] \rightsquigarrow \& x \dashv \Gamma
    }

    \inferrule[E-Array-Slice]{
      x: [T; N] \in \Gamma
    }{
      \Gamma \vdash x \triangleleft \&[T] \rightsquigarrow \& x[..] \dashv \Gamma
    }

    \inferrule[E-Slice-Box]{
      x: \&[T] \in \Gamma \\
      \Gamma' = \Gamma \setminus \{ x \}
    }{
      \Gamma \vdash x \triangleleft \mathsf{Box}\langle[T]\rangle \rightsquigarrow
      (*x).\mathsf{into}() \dashv \Gamma'
    }

    \inferrule[E-Array-Box]{
      x: [T; N] \in \Gamma \\
      \Gamma' = \Gamma \setminus \{ x \}
    }{
      \Gamma \vdash x \triangleleft \mathsf{Box}\langle[T]\rangle \rightsquigarrow
      \mathsf{Box}::\mathsf{new}(x) \dashv \Gamma'
    }

    \inferrule[E-AddrOf]{
      \Gamma \vdash e \triangleleft [T; N] \rightsquigarrow E \dashv \Gamma'
    }{
      \Gamma \vdash \& e \triangleleft \&[T] \rightsquigarrow \&E \dashv \Gamma'
    }

    \inferrule[E-Call]{
      f: (\vec T) \to T \in \Gamma \\
      \Gamma_i \vdash e_i \triangleleft T_i \rightsquigarrow E_i \dashv \Gamma_{i+1} \\\\
      \Gamma_n, r: T \vdash r \triangleleft T' \rightsquigarrow E' \dashv \Gamma'
    }{
      \Gamma_0 \vdash f(\vec e) \triangleleft T' \rightsquigarrow \mathsf{let}\ r = f(\vec E);\; E' \dashv \Gamma' \setminus \{ r \}
    }

  \end{mathpar}
  \caption{Translation from \minic to Rust, Selected Rules}
  \label{fig:expressions}
\end{figure*}

\subsection{Compiling Pointer Arithmetic}
\label{sec:split}

We now reach the main difficulty of the translation, namely, handling \emph{pointer arithmetic}.
When operating on arrays, C programs rarely perform accesses and updates via a single
base pointer. A common pattern is to instead divide the array into chunks,
or iterate over the array elements by keeping a local pointer to the current head of
the iterator.

Consider the following C example (\fref{splitcode}, left),
inspired by an implementation of elliptic-curve cryptography.
The array \li{abcd} contains a large number (``bignum''), spread across four
64-bit (8 bytes) \emph{limbs} stored contiguously in memory. For field addition,
a first order of business is to get pointers to individual limbs, before
performing pointwise addition. In other words,
we want to perform pointer arithmetic to access chunks \li{a}, \li{b}, \li{c},
\li{d}, spanning intervals $[0; 8)$, $[8; 16)$, $[16; 24)$ and $[24; 32)$ of the
base pointer.
Because C does not carry length information for pointers, neither at run-time nor in
the type system, we do not know that each pointer intends to span 8 bytes.
Furthermore, we cannot assume either that pointer arithmetic occurs in
left-to-right order: \fref{splitcode} illustrates this.

\begin{figure*}
  \begin{minipage}{0.48\linewidth-2em}%
    \begin{minted}[xleftmargin=0.5cm, resetmargins]{c}
uint8_t abcd[32] = { 0 };

uint8_t *a = abcd + 0;
uint8_t *c = abcd + 16;
uint8_t *b = abcd + 8;
uint8_t *d = abcd + 24;

    \end{minted}
  \end{minipage}%
  \hspace{2em}\vrule\hspace{2em}%
  \begin{minipage}{0.48\linewidth-2em}%
    \begin{minted}[resetmargins]{rust}
let mut abcd = [0u8; 32];
let abcd: &mut [u8] = &mut abcd[..];
let (a_l, a_r) = abcd.split_at_mut(0);
let (c_l, c_r) = a_r.split_at_mut(16);
let (b_l, b_r) = c_l.split_at_mut(8);
let (d_l, d_r) = c_r.split_at_mut(8);
let (a, b, c, d) = (b_l, b_r, d_l, d_r)
    \end{minted}
  \end{minipage}%
  \caption{Example of pointer arithmetic translation as Rust splits. For readability, we destructure tuples instead of using tuple variables and accessors as described in \fref{algo}.}
  \label{fig:splitcode}
\end{figure*}

This bit of C code cannot be trivially translated to Rust, because
in order to guarantee memory safety, Rust does not allow arbitrary pointer arithmetic.
What Rust provides instead is a primitive named \li{split_at_mut} (or \li{split_at} for immutable slices), which
allows the programmer to relinquish ownership of a slice, and obtain in exchange two sub slices that
split the original slice at the given index.
This permits some restricted
notion of pointer arithmetic, while preserving Rust's invariant that mutable data should have a
unique owner: to regain ownership of the original slice, the programmer must give up the sub slices.
To reconcile C's pointer arithmetic with Rust's splitting paradigm, we therefore propose a novel
notion of \emph{split trees}.

\paragraph{Split Trees}

\begin{figure}[t]
\small
    \begin{algorithmic}
\LComment{When a new variable x enters the scope other
than through \texttt{let x = base + index} (function parameter, stack or heap
allocations, etc.): update environment}
\Procedure{init}{x}
  \State Tree(x) $\gets$ Leaf
  \State Base(x) $\gets$ None
  \State Path(x) $\gets$ []
\EndProcedure

\vspace{1ex}

\LComment{When compiling \texttt{let x = base + index}:}
\Function{new-split}{x, base, index}
  \LComment{Locate the position where we need to split in the binary search tree for base}
  \State path, node $\gets$ \Call{Binary-Search}{Tree(base), index}

  \LComment{Extend the split tree of base: a new sub-child is added to node.}
  \If{index < node.index}
    \State suffix $\gets$ Left
  \Else
    \State suffix $\gets$ Right
  \EndIf
  \State Tree(base)[node].suffix $\gets$ \Call{Node}{index, Leaf, Leaf}

  \LComment{Fill out the environment:}
  \State Tree(x) $\gets$ Leaf \Comment{x itself has not been split}
  \State Base(x) $\gets$ Some base \Comment{x now holds a sub-range of base, described in the split tree of base...}
  \State Path(x) $\gets$ path ++ [ suffix ] \Comment{... by the node found at the end of path, followed by suffix}

  \LComment{To produce syntax for x, we need y, the variable in scope
  for its parent node. Note that y has a pair type.}
  \State y $\gets$ \Call{find}{} y \Call{in scope such that}{} Base(y) = Some base \textbf{and} Path(y) = path

  \LComment{Moving right requires subtractions -- see e.g. \texttt{c\_r} in
\fref{tree-final}. We encapsulate this in \Call{Compute}{} (elided)}
  \State offset $\gets$ \Call{compute}{Tree(base), path, index}

  \LComment{Return a Rust expression. \Call{field-of-suffix}{} is 0 for Left, 1 for
  Right -- this is Rust syntax for pair projections.}
  \State \Return \texttt{let x = y.}\Call{field-of-suffix}{\texttt{suffix}}\texttt{.split\_at(offset);}
  \Comment{Note that x is now a pair.}

  \EndFunction

\vspace{1ex}

\LComment{When compiling \texttt{x} (reference to variable \texttt{x})}
\Function{var-lookup}{x}
  \If{Base(x) = None}
      \State \Return \texttt{x} \Comment{This variable has not been subjected to splits
      -- trivial compilation scheme}
  \ElsIf{Base(x) = Some base}
      \State y, suffix $\gets$
        \Call{find nearest}{} y, suffix \Call{in scope such that}{}
      \State \phantom{y, suffix $\gets$}
          Base(y) = base \textbf{and}
          Path(x) \Call{is-accessible-via}{} Path(y) = Some suffix
      \State \Return \texttt{y.}\Call{field-of-suffix}{\texttt{suffix}}
  \EndIf
  \EndFunction

\vspace{1ex}

\Function{is-accessible-via}{p1, p2}
  \If{p2 = p1}
    \LComment{If no further split occurred, elements of interest are on the
    right}
    \State \Return Some Right \Comment{e.g., \texttt{a} is accessible via \texttt{a\_r} in \fref{tree-step1}}
  \ElsIf{\Call{exists}{} n >= 0 \Call{such that}{} p2 = p1 ++ [ Right ] ++ \Call{repeat}{Left, n}}
    \LComment{If a split occurred, move right (to get the elements of interest),
    then keep left (to avoid other variables)}
    \State \Return Some Left \Comment{
      e.g., \texttt{a} is accessible via \texttt{c\_r} in \fref{tree-step2}, and
      via \texttt{d\_l} in \fref{tree-final}
    }
  \Else
    \State \Return None
  \EndIf
\EndFunction

    \end{algorithmic}

  \caption{An algorithm for reconstructing pointer arithmetic using split trees;
[] is syntax for the empty list.}
  \label{fig:algo}
\end{figure}

Our compilation algorithm is described in \fref{algo}; we use \fref{splitcode}
as a supporting example to illustrate how it works.
When translating C's pointer arithmetic to Rust, several difficulties arise.
First, because we do not have length information coming from the C side, we need to
assume that the chunks are not intended to be overlapping -- if they were, this
code would simply be impossible to type-check, and the programmer would have to rewrite their C code
to make the intent more apparent, keeping in line with our semi-active approach to translating C to
Rust.
Second, the translation needs to be predictable and understandable by the user, so that translation
failures can easily be matched with the location in the original C code that needs to be rewritten.
For those reasons, we want to avoid backtracking in our translation, and perform the translation in a forward fashion.
This means that we need a data structure that keeps
the history of the calls to \li{split_at}, for instance to know at line 5 that C index 8 lives in
\li{c_l}, and at line 6 that C index 24 lives in \li{c_r}, at Rust index $24-16=8$.
This data structure also must be attached to every program point,
so as to translate an access through a C pointer into an access (possibly with
an offset computation) through a Rust slice. For instance, \li{a[0]} would translate into
\li{a_r[0]} after line 3, but into \li{c_l[0]} after line 4, and so on.

We solve these challenges with a data structure called a split tree, synthesized during our
translation: each C pointer maps to a (possibly singleton) split tree, which evolves in a flow-dependent
fashion. We present the split trees corresponding to \li{abcd} at different points of our translation
in Figure~\ref{fig:splittree}; the split tree initially only contains the root \li{abcd}.

\begin{figure}[h]
  \begin{subfigure}[t]{0.20\textwidth}
    \centering
    \begin{tikzpicture}[level distance=8mm,level/.style={sibling distance=18mm/#1}]
      \node {(abcd, 0)}
      child {node {a\_l} edge from parent node [left,yshift=3pt] {}}
      child {node {a\_r} edge from parent node [right,yshift=3pt] {}};
    \end{tikzpicture}
    \caption{After translating a}
    \label{fig:tree-step1}
  \end{subfigure}
  \begin{subfigure}{0.5\textwidth}

  \end{subfigure}
  \begin{subfigure}[t]{0.25\textwidth}
    \centering
    \begin{tikzpicture}[level distance=8mm,level/.style={sibling distance=18mm/#1}]
      \node {(abcd, 0)}
      child {node {a\_l} edge from parent node [left,yshift=3pt] {} }
      child { node  {(a\_r, 16)}
        child {node [xshift=-0.3cm] {c\_l} edge from parent node [left, yshift=3pt] {} }
        child {node [xshift=0.3cm] {c\_r} edge from parent node [right, yshift=3pt] {} }
        edge from parent node [right,yshift=3pt] {}
      };
    \end{tikzpicture}
    \caption{After translating c}
    \label{fig:tree-step2}
  \end{subfigure}
  \begin{subfigure}[t]{0.40\textwidth}
    \centering
    \begin{tikzpicture}[level distance=8mm,level/.style={sibling distance=28mm/#1}]
      \node {(abcd, 0)}
      child {node {a\_l} edge from parent node [left,yshift=3pt] {}}
      child {node {(a\_r, 16)}
        child {node [xshift=-0.3cm] {(c\_l, 8)}
          child {node {b\_l} edge from parent node [left, yshift=3pt] {} }
          child {node {b\_r} edge from parent node [right, yshift=3pt] {} }
          edge from parent node [left, yshift=3pt] {}
        }
        child {node [xshift=0.3cm] {(c\_r, 8)}
          child {node {d\_l} edge from parent node [left, yshift=3pt] {} }
          child {node {d\_r} edge from parent node [right, yshift=3pt] {} }
          edge from parent node [right, yshift=3pt] {}
        }
        edge from parent node [right,yshift=3pt] {}
      };
    \end{tikzpicture}
    \caption{After translating b and d}
    \label{fig:tree-final}
  \end{subfigure}

  \caption{Successive split trees during C translation. Internal nodes of the form (x, i) have been
  subjected to a split at Rust index i and are therefore borrowed at this program point. Leaf nodes are
  available.}
  \label{fig:splittree}
\end{figure}

The first pointer addition defines \li{a} as a sub-array of \li{abcd} starting at index 0, whose
intended length is unknown. We thus split \li{abcd} at index 0, and keep this information, meaning
indices $[0;0)$ of \li{abcd} are in the left slice, \li{a_l}, and indices $[0; 32)$ are in \li{a_r} (Figure~\ref{fig:tree-step1}).
The second pointer addition on the same base pointer \li{abcd} triggers another split. At this
program point, \li{abcd} is no longer available, because it has been \emph{borrowed} to construct
\li{(a_l, a_r)}. Splitting \li{abcd} directly would be a mistake, because it would terminate these
borrows, and render \li{(a_l, a_r)} unusable, thus making it impossible to translate any further
usage of C pointer \li{a}. Instead, we leverage the fact that our split tree is a binary search
tree, and discover that key \li{16} needs to be inserted as the right child of node \li{a}, i.e.,
we must split \li{a_r} at index 16. At this stage, a \emph{use} of C pointer \li{a} would trigger a
search in the binary tree, and would return \li{c_l} as the current slice through which
\li{a} may be accessed (Figure~\ref{fig:tree-step2}). The translation of \li{b} is similar. Finally, for \li{d}, we find that
index \li{24} is to be found in \li{c_r}; because the indices of the right subslice restart from 0,
we must perform a subtraction to know that \li{c_r} needs to be split at index 8 (Figure~\ref{fig:tree-final}).

We generalize this mechanism to all variables in the environment, and equip it with a few more bells
and whistles (not depicted in \fref{algo}). First, any usage of \li{abcd} that is not pointer arithmetic (e.g., \li{f(abcd)}) is taken to indicate that
C variables \li{a}, \li{b}, \li{c}, and \li{d} are no longer useful and that the user intends to
perform a fresh set of pointer arithmetic computations.
This allows the programmer to insert, in the
source code, calls to, e.g., \mintinline{c}{(void)abcd} to provide a hint that the split tree of \li{abcd} needs
to be reset, and that the corresponding Rust variables can go out of scope (i.e., their lifetimes can
end). Such calls are optimized away by the compiler, and therefore do not have any impact at runtime.
We leveraged this mechanism in numerous places in our \haclstar case study
(Section~\ref{sec:hacl}). For instance, in elliptic curve implementations,
it is common to have a function be a series of calls to \li{ADD} and \li{MUL} macros,
each of which expands to operations that need their own split trees.
Second, we
generalize the form of offsets to accept a more general language of expressions,
described below.

\paragraph{Symbolic Solver}

The split trees we presented above operate over constant offsets -- this
allows implementing a trivial order, which is required for the binary search tree
structure. In real-world examples, we commonly encountered more complex offset
expressions, for instance \li{n} and \li{n + 8} where \li{n} might be a parameter
of the current function, whose value is therefore not statically known.

To address this issue, our implementation relies on a simple, deterministic symbolic solver,
which is able to compare different kinds of arithmetic expressions containing
symbolic variables, e.g., to determine that \li{n + 8} is greater than \li{n}.
While this solver is not complete and does not rely on contextual information
to refine the analysis, it is sufficient to translate large case
studies as described in Section~\ref{sec:casestudies}, and could be easily
extended for further use-cases.

Should the symbolic solver still fail to compare offsets, our implementation emits a
warning, along with the corresponding location. This allowed us to fix the source
code in our case studies, for instance, to replace \li{2*n} with \li{n+n}. Should none of this work,
our compiler then adopts a final heuristic: that the offsets that occur along
the control-flow are monotonically increasing, i.e., they perform pointer
arithmetic
from left to right.
Again, our semi-active translation approach means that, in case
the programmer's intent is unclear, it is oftentimes worthwhile to rewrite the source, rather than
augment our solver with complex heuristics.

\paragraph{Limitations}

We now comment on this approach. If there is true aliasing, i.e., the C program
contains both \li{a[8]} and \li{b[0]} in our earlier example, the translated Rust code will perform
an out-of-bounds array access.
In other words, we must assume the array chunks, in C, to be disjoint.
For overlap cases that can be distinguished statically (as above), we emit a compile-time error; otherwise, the
Rust code will panic at runtime.
We remark that we have not seen this
pattern in our case studies, so this has not been a problem in our experience.
Should this turn out to be a concern in the future, we envision either supplemental off-the-shelf
static analyses to determine slice lengths, or additional annotations in the source code to express intent more clearly.

\subsection{Mutability Inference}
\label{sec:mutability}
To ensure memory safety, Rust relies on a principle called mutability XOR
aliasing. Concretely, it distinguishes between immutable borrows, of type
\li{&T}, which can be freely shared and duplicated but only allow to read the
referenced memory, and mutable borrows, of type \li{&mut T}, which allow to
write memory, but require exclusive ownership, that is, no other reference to
the same memory region can be live at the same time. This is enforced by the Rust
\emph{borrow-checker}.

The distinction between mutable and immutable borrows raises two conflicting problems.
To pass borrow-checking, one needs to provide mutability annotations,
to ensure that memory being modified is indeed mutably borrowed.
Unfortunately, marking all borrows as mutable is not a solution. Beyond being unidiomatic,
a function requiring only mutable borrows is also restrictive,
as it prevents usages where aliasing would be safe, i.e., when arguments are only
read.

To reach a middle ground, our Rust translation
generates immutable borrows by default, which we now refine as-needed into
mutable borrows through a custom mutability inference analysis. To do so,
we perform a backward analysis on all translated functions, identifying expressions
that perform memory updates, and backpropagating which program variables need
to be mutable back to their definition site, inserting mutable borrows along the way.

Mutability rules are as follows in Rust.
If \li{x} has array type \li{[T; N]} and \li{x[i] = e} occurs, then
\li{x} must be a mutable variable, i.e., declared as \li{let mut x = ...}.
If \li{y} has borrow type and \li{y[i] = e} occurs, then \li{y} must be a
\emph{mutable} borrow; furthermore, one can only mutably borrow
variables that are themselves mutable, i.e., if \li{let y: &mut [u8] = &mut z},
then \li{z} itself must be declared as \li{let mut z}.

We inductively define an analysis on the Rust syntax that is aware of the
semantics above and synthesizes two variable sets $V$ and $R$, where $V$ contains
variables that must be mutable (i.e., \li{x} and \li{z}, above), and $R$ contains variables
that must have a mutable borrow \emph{type} (i.e., \li{y} above). Applying
this analysis to the output of our translation yields a Rust program that has been annotated with the minimum amount
of \li{mut} qualifiers in variables,
types, and function parameters, in order to type-check. In practice, our
analysis is more general, and can handle many more forms of assignments,
with combinations of fields, borrows, array indexing, and so on. Our
implementation also computes a third set $F$, for fields of structs that ought to
be mutable, and a fourth set $P$, for pattern variables in \li{match}es that ought
to be \li{ref mut} -- we omit these details here for the sake of simplicity.

We formally present our mutability inference analysis in
Figure~\ref{fig:mutability}. Our rules are presented as the combination of a judgment
$\Delta \vdash e \mutrel{V, R}{M} e', V', R'$, and a system of mutually-recursive equations over
$\Delta$. The final output of our analysis is the least fixed point that satisfies the equations
over $\Delta$.
The $\mutrel{V,R}M$ judgment represents that Rust expression $e$ is
transformed into expression $e'$. This translation is performed with current
sets $V$, $R$, and returns new sets $V'$ and $R'$, as the translation of $e$ might have
added variables to both sets. The mode $M \in \{mut, imm\}$ indicates the
expected borrow mutability of the expression $e$, while $\Delta$ contains the function
definitions, which are needed to retrieve the expected mutability when
performing function calls.

\begin{figure}[h]
  \centering
  \small
  \begin{mathpar}
    \inferrule[I-ImmVar]{ }{
      \Delta \vdash x \mutrel{V,R}{imm} x, V, R
    }

    \inferrule[I-MutVar]{ }{
      \Delta \vdash x \mutrel{V,R}{mut} x, V, R \cup \{ x \}
    }

    \inferrule[I-ImmBorrow]{ }{
      \Delta \vdash \&x \mutrel{V,R}{imm} \&x, V, R
    }

    \inferrule[I-MutBorrow]{ }{
      \Delta \vdash \&x \mutrel{V,R}{mut} \&mut~x, V \cup \{x\}, R
    }

    \inferrule[I-Let]{
      \Delta \vdash e_2 \mutrel{V,R}{M} e'_2, V', R' \\
      M' = \ite{x \in R'}{mut}{imm} \\
      A = \ite{x \in V'}{mut}{\emptyset} \\
      \Delta \vdash e_1 \mutrel{V', R'}{M'} e'_1, V'', R''
    }{
      \Delta \vdash \mathsf{let}\; x = e_1; e_2 \mutrel{V,R}{M} \mathsf{let}\; A\; x = e'_1; e'_2, V'', R''
    }

    \inferrule[I-Update-Borrow]{
      \Delta \vdash e_1 \mutrel{V, R}{mut} e'_1, V', R' \\
      \Delta \vdash e_2 \mutrel{V, R}{imm} e'_2, V'', R'' \\
      \Delta \vdash e_3 \mutrel{V, R}{imm} e'_3, V''', R''' \\
    }{
      \Delta \vdash e_1[e_2] = e_3 \mutrel{V,R}{M} e'_1[e'_2] = e'_3, V' \cup V'' \cup V''', R' \cup R'' \cup R'''
    }

    \inferrule[I-Call]{
      \Delta(f) = (x_1: T_1, \ldots x_n: T_n) \to T \\
      \forall i, \Delta \vdash e_i \mutrel{V,R}{is\_mutborrow(T_i)} e'_i, V_i, R_i \\
      T' = \text{if } M = mut \text{ then } make\_mut(T)\text{ else } T
    }{
      \Delta \vdash f(e_1, \ldots, e_n) \mutrel{V, R}{M} f(e'_1, \ldots, e'_n), \bigcup_i V_i, \bigcup_i R_i \\\\
      \Delta(f) = (x_1: T_1, \ldots x_n: T_n) \to T'
    }

    \inferrule[I-Sig]{
      R = \left\{ x_i \mid is\_mutborrow(T_i) \right\} \\
      \Delta \vdash e \mutrel{\emptyset, R}{is\_mutborrow(T)} e', V', R' \\
      \forall i, T'_i = \ite{x_i \in R'}{make\_mut(T_i)}{T_i} \\
      \Delta(f) = (x_1: T_1, \ldots, x_n: T_n) \rightarrow T \{ e \}
    }{
      \Delta(f) = (x_1: T'_1, \ldots, x_n: T'_n) \rightarrow T \{ e' \}
    }
  \end{mathpar}
  \caption{Mutability Inference Analysis, Selected Rules}
  \label{fig:mutability}
\end{figure}

Rules \Rule{I-ImmVar} and \Rule{I-MutVar} are straightforward:
if the translation expects variable $x$ to be a mutable borrow, then
we add it to the set $R$ to backpropagate the information, otherwise,
we leave both sets invariant.
Rules \Rule{I-ImmBorrow} and \Rule{I-MutBorrow} are similar, but
operate directly on borrows: if variable $x$ is borrowed and the
expected type is a mutable borrow, then we
replace the immutable borrow by a mutable borrow, and indicate
that variable $x$ must be mutable.

Rule \Rule{I-Let} demonstrates the backward nature of the analysis.
As the type of the expression
corresponds to the type of $e_2$, we first translate $e_2$ with
the same mode $M$, returning new sets $V'$ and $R'$. We then rely
on these sets to translate $e_1$. If $x$ belongs to
$R'$, then it means that the rest of the program expects
it to have a borrow type, we thus translate its definition with
mode $mut$. Finally, if $x$ is used mutably, i.e., belongs to $V'$, we make its let-binding
\li+mut+.

Rule \Rule{I-Update-Borrow} presents the translation of a
borrowed slice update, which introduces the need for mutable borrows.
Our translation from C to \minic guarantees
that the left-hand side has a pointer type (\Rule{Index}) which is then translated
to a Rust slice borrow type.
To satisfy the borrow-checker, the expression $e_1$
being modified must thus become a \emph{mutable} slice borrow.
We therefore
translate with the $mut$ mode. Conversely, both the index
and the value being stored are only read; they are
translated with mode $imm$.

The translation of function calls is handled through rule
\Rule{I-Call}. When calling a function $f$, we first retrieve
the type signature of $f$ from the environment $\Delta$.
We then translate the expressions corresponding to each function
argument according to their expected mutability, computed
through the function $is\_mutborrow(T_i)$. This function
returns $mut$ if $T_i$ is of the shape \li{&mut T}, and $imm$
in all other cases.
The translation of all arguments can be done in parallel with
the initial sets $V$, $R$. To propagate the information acquired,
we finally return the union of these sets; this faithfully
models that if an expression $e_i$ requires a variable $x$
to be mutable while $e_j$ does not have this requirement,
then $x$ must indeed be mutable. A key feature of this rule is that it may modify the definition
environment $\Delta$, and thus trigger further recomputations to reach the fixed-point of our system
of equations. Indeed, the context may force us to produce a mutable borrow, which can only be
achieved by forcing the return type of \li+f+ itself to be a mutable borrow.

To infer mutability information for an entire program, our analysis
operates over top-level function definitions in
\Rule{I-Sig}. This rule can be seen as the entrypoint of our analysis.
Given a previous iteration of the analysis for $f$,
we compute the initial set
$R$ of variables whose type is a mutable borrow,
based on the signature
of $f$. The translation of function body $e$ returns a new function body $e'$, as well
as sets $V'$ and $R'$. We update the definitions $\Delta$ with a new entry for $f$,
based on information inferred during the translation of $e$: if a function
parameter $x_i$ belongs to the set $R'$, meaning it is expected to
be a mutable borrow, then we modify its type $T_i$ accordingly.

The presentation as a system of iterated equations alleviates the need for a topological sort of the
function definitions -- there may simply be no such order, given mutually-recursive definitions in C
and Rust. The environment is initially populated with the output of our translation, meaning there
are no \li{mut} qualifiers anywhere. We then iterate the equations over
$\Delta$ until a fixed point is reached. There trivially exists such a fixed point: we only ever add
\li{mut} qualifiers, meaning the number of iterations is bounded by the numbers of function
parameters across the whole program.
In practice, we of course do not do repeated iterations, and instead use an optimized fixed point
library to eliminate un-necessary recomputations~\cite{pottier2009lazy}.

While Figure~\ref{fig:mutability} presents the essence of our mutability
inference algorithm, as mentioned earlier, our implementation supports a
much larger set of features, to translate our real-world case studies
(Section~\ref{sec:casestudies}).
Importantly, this analysis does not modify the Rust program; it only augments it
with additional \li{mut} qualifiers, meaning it can be validated \emph{a posteriori}
by the Rust compiler.

\paragraph{{Mutability in Structs and Tuples}}
We mentioned earlier that compiling some C structs to Rust tuples was useful not just for code
quality, but also for accepting more programs. We now explain why. Consider a C program, equipped
with a struct type \li+t+ that holds two integer pointers, and two functions
\li+t f(uint32_t *src)+ and \li+t f_mut(uint32_t *src)+. If \li+t+ compiles to a Rust struct,
and if \li+f_mut+ returns pointers that are found to be \li+&mut+ by our analysis, then
our analysis will propagate this information and make the two fields of \li+t+ have type
\li+&'a mut u32+; this, in turn, will force \li+f+ to return mutable pointers too.

If instead \li+t+ compiles to a Rust tuple, we can have both \li+fn f(src: &[u32]) -> (&[u32], &[u32])+
and \li+fn f_mut(src: &mut[u32]) -> (&mut[u32], &mut[u32])+. Because the two tuple types are
unrelated (we say that they obey structural typing, as opposed to nominal typing like structs),
making one mutable does not affect the other and therefore avoids type-checking errors in
\li+f+.

\paragraph{{Support for slices}}
Some C programs already manipulate the equivalent of a slice representation; indeed, it is not
uncommon, in security-sensitive contexts, to carry pointers along with their lengths. Our
implementation supports special annotations that designate such structs as officially
representing a slice, meaning occurrences of this struct compile directly to a Rust slice.

\subsection{Automatically Deriving Traits Instances}
\label{sec:traits}

A key component of the Rust language is its pervasive use of \emph{traits},
which can be broadly seen as a Rust-specific version
of typeclasses~\cite{wadler1989typeclasses}. Traits
allow programmers to specify that a given type must implement certain features,
for instance that it can be pretty-printed (\li{Display} trait), converted from
and to another type (\li{From} and \li{Into} traits), or that it allows deep
copies (\li{Clone} trait).

When defining a new type \li{T}, i.e., a new structure or enumeration, one can
provide an \emph{implementation} of a given trait $\mathfrak{T}$ by defining
the methods corresponding to $\mathfrak{T}$ for type \li{T}, e.g., implementing a
\li{fmt} function that specifies how to pretty-print an element of type \li{T}
allows to implement trait \li{Display} for \li{T}.  For some traits however,
implementations can be automatically derived, using the Rust
\li{#[derive(Trait)]} attribute on the type definition, assuming that all
fields of the structure or enumeration do implement the trait.

One trait of particular interest for our translation is the \li{Copy} trait. A
common pattern in C is to define a structure containing a pointer (e.g., a
pointer to a string and the corresponding length), and to pass the structure by
value when calling functions.  This leads to a mismatch with the Rust move
semantics.  Rust implements an affine type system, meaning that, by default,
values can be used at most once: passing a value to a function, e.g.,
\li{f(s)}, invalidates the value $s$, leading to compile-time errors if \li{s}
is further used in the program.  To avoid this behavior, types can implement
the \li{Copy} trait, which instead performs a shallow copy of the value when
passing it to a function.

To avoid Rust compilation errors,
we thus wish to automatically derive \li{Copy}
when possible. All basic Rust types (e.g., integers or booleans) and immutable borrow types \li{&T}
implement \li{Copy}.
However, copying a mutable pointer
(i.e., either \li{&mut T} or \li{Box<T>}) is not allowed; this would contradict
Rust's guarantee that every piece of mutable data has a unique owner.  To automatically add a
\li{#[derive(Copy)]} annotation on appropriate structure and enumeration type
definitions, we traverse all type definitions, and only derive the \li{Copy}
trait if all fields are themselves \li{Copy}, that is, they are neither a
mutable borrow nor a box, and if they are a custom type, this type implements
\li{Copy} itself. The analysis is also performed through a fixpoint
computation to handle (mutually) recursive structs and enums.

In addition to \li{Copy}, our analysis also automatically derives
the \li{PartialEq} (allowing the use of the \li{==} operator), and
\li{Clone} traits when possible. \li{Clone}
allows performing deep copies on a given value, via an explicit call to the \li+clone+ method.
While our translation does not directly use this feature, implementing \li{Clone} is a
prerequisite to implement \li{Copy} (in Rust's parlance, \li{Copy} has \li{Clone} as
a parent trait).
Additionally, \li{Clone} is commonly used by library clients; we therefore strive to
provide it as much as possible to facilitate the adoption of the translated Rust library.

\section{Evaluation}
\label{sec:casestudies}

\subsection{Implementation}

We implement the compilation strategy described in this paper in a tool called Scylla. Scylla is
written in about ~7,200 lines of OCaml, and took 1.5 person-year to author. For parsing actual C code,
Scylla relies on OCaml's \li+libclang+ bindings~\cite{clangml};
for compilation to Rust, Scylla relies on a novel
Rust representation in OCaml, complete with printer, visitors for optimization passes, and the
implementation of the various analyses and fixed-point computations we describe in this paper.
Finally, for \minic, Scylla relies on the existing KaRaMeL~\cite{lowstar} project.

Concerning the scope of modifications that were required for various projects to
translate to safe Rust, we estimate that no more than a few days of work were
spent on each project. This also includes fixing the proofs in \haclstar.
For all projects, we confirmed that the Scylla-translated Rust code compiled and
passed tests that we manually ported.

\begin{table}
  \small
  \caption{Performance comparison. Results are normalized taking the original C code as a baseline.}
  \begin{tabular}{llll}
    \hline
    Benchmark & Original C & Rewritten C & Rust \\
    \hline
    SymCrypt SHA3-256       & 1 & 1.00 & 1.00 \\
    SymCrypt SHA3-384       & 1 & 1.00 & 1.01 \\
    SymCrypt SHA3-512       & 1 & 1.00 & 1.00 \\
    Frodo640KEM KeyGen      & 1 & .98 & .97 \\
    Frodo640KEM Encaps      & 1 & .99 & .98 \\
    Frodo640KEM Decaps      & 1 & .99 & .99 \\
    \haclstar Curve25519    & 1 & 1.02 & .99 \\
    \haclstar SHA2-256      & 1 & .99 & 1.05 \\
    \haclstar ChachaPoly-Enc& 1 & 1.00 & .99 \\
    \haclstar ECDH-P256     & 1 & .98 & .77 \\
    bzip2-compress          & 1 & .99 & 1.08 \\
    EverParse CBOR 2200     & 1 & 1.00 & .87 \\
    EverParse CBOR 1000     & 1 & 1.00 & 1.21 \\
    EverParse CBOR 250      & 1 & 1.00 & 1.22 \\
    EverParse CBOR 2200 (inline) & 1 & 1.00 & .65 \\
    EverParse CBOR 1000 (inline) & 1 & 1.00 & .91 \\
    EverParse CBOR 250  (inline) & 1 & 1.00 & .91 \\
  \end{tabular}
  \label{table:perf}
  \vspace{-0.5cm}
\end{table}

\subsection{Case Studies}

We evaluate our methodology on five real-world codebases that exemplify the kind of
regular, applicative C code that we can convert to safe Rust. For each of those, we first
describe the changes we had to enact to make the code eligible for translation.
Then, we measure performance, quantifying both the
cost of restructuring the C code to fit \minic,
and the performance impact of switching to Rust. When comparing C
and Rust, we use the \texttt{clang} and \texttt{rustc} compilers, and
make sure that the version of LLVM is identical for both.

All experiments were run on a MacOS 15.5 machine with an ARM M3 Max processor and
96GB of RAM. Aiming for an apples to apples comparison,
we compile the C code with the \li{-O2} optimization level,
while the Rust code is compiled in release mode, with \li{opt-level} set to 2.
To minimize the impact of differences in compilation toolchains, e.g., Rust
treating an entire crate as a single translation unit, which opens up
tremendous optimization opportunities, we enable link-time
optimization (LTO) in both C and Rust. In C, we do this by passing the
\li{-flto} option to \li{clang}, while we set \li{lto = "fat"} in Rust.

\subsubsection{SymCrypt}

SymCrypt is the cryptographic library used in all Microsoft products and services. It is written in
C, and for a few years now, has been open-source~\cite{symcrypt}.
We ran Scylla on its Keccak implementation, which
includes four variants of SHA3~\cite{sha3-standard} and two variants of SHAKE~\cite{sha3-standard} and CSHAKE~\cite{cshake-standard}. After marking SymCrypt-specific
helpers for endianness conversion as opaque (a one-line change for about twenty helpers), the code
was converted to safe Rust out of the box. For code quality, we replaced C macros with
\li+static inline+ variants, which produced functions in Rust, instead of translating the result of
macro-expansion.
Out of 1814 lines of C code, the diff utility reports 91 insertions and
61 deletions for the Keccak-related files.
22 insertions correspond to the straightforward addition of Scylla opaque annotations.
The rest corresponds to the replacement of 13 C macros by \li`static inline` functions.
While each line of the macros needs to be modified due to the C macro syntax requiring trailing \li`\` characters compared to function bodies, therefore increasing the number of lines modified,
changes were completely mechanical.

For performance benchmarking, we added FFI bindings to our Rust Keccak implementation that had the
exact same API as the original C code, down to the layout of the structs. We were then able to use
SymCrypt's internal benchmarking utility to measure either the original C implementation, or our own
Rust implementation exposed through the same C ABI. Because we enacted no structural changes on the
source code, the performance is identical.

\subsubsection{FrodoKEM}

We took the reference Microsoft C implementation of FrodoKEM~\cite{frodokem}, a post-quantum key-exchange mechanism~\cite{frodokem-rfc}, and set
out to translate it to Rust. This was perhaps our most interesting example.
First, we adopted a hybrid approach. FrodoKEM relies on two primitives:
AES~\cite{aes-standard} and SHA3~\cite{sha3-standard}. We chose to
translate SHA3, but not AES, because none of the AES variants packaged in FrodoKEM were eligible for
translation. The pure C variant violates strict aliasing, meaning it not only exercises undefined
behavior, but is also ineligible for translation to Scylla. The version that leverages hardware
instructions is not supported by Scylla either, as we do not yet compile compiler intrinsics to their
Rust equivalents. We simply created a hybrid build, where AES remains in C, and we wrote FFI
bindings to allow our translated Rust code to call the original AES. For a real-world
scenario, we would simply ``fill out'' the AES implementation using an existing crate.
The second reason this example is of particular interest is that it uncovered
several cases of undefined behavior (UB) in the FrodoKEM implementation. Cryptographic code oftentimes
converts between \li+char*+ and \li+uintN_t*+, which is legal in C, and which Scylla supports, using
a mechanism similar to the \li+zerocopy+ crate~\cite{zerocopy}.
However, compared to C, Rust checks such conversions for proper alignment in debug builds.
As it turns out, alignment was violated in
Rust, which we traced back to an alignment violation in C, which is UB.
Furthermore, we also uncovered a conversion between \li+uint16_t*+ and \li+uint32_t*+, which we do
not support in Scylla, because it is UB in C. Both UBs were
responsibly disclosed and reported upstream, along with a suggested fix.

Beyond that, the bulk of our changes revolved around lifetimes: the code was written C89-style, with
all variables declared at the top of the function, which generated un-necessary long lifetimes in
Rust. We solved these by moving variable declarations closer to their use-site, or by using a macro to
replace a local variable with its definition when it
is trivial, e.g. turn \li!int *x = y + 3! into \li!#define x (y + 3)!.
Out of 1451 lines of C code, the diff utility reports 5.5\% of the source code impacted. %
Most of these modifications were highly mechanical: moving variable declarations represents 75\% of these changes, while
copying small arrays passed as function arguments to satisfy the Rust borrow checker represents an additional 17\%.

The Rust code is slightly faster than the original C; the illegal cast from
\li{uint16_t*} to \li{uint32_t*} was intended to implement a \li{memcpy}, 4 bytes
at a time. Our fix was to replace it with an actual \li{memcpy} instruction,
which probably allowed the compiler to optimize a little bit more.

\subsubsection{\haclstar}
\label{sec:hacl}
\haclstar is a verified cryptographic library~\cite{haclxn},
compiled to C from \lowstar~\cite{lowstar}, a low-level dialect of
the \fstar programming language~\cite{mumon}.
Because \lowstar contains very few constructs, namely, arrays, machine
integers, structures, pointer arithmetic, and primitive array operations such as \li+blit+ or
\li+fill+, the resulting C code is very regular, and is a prime candidate for translation to safe
Rust.
As part of our evaluation, we had to rewrite some code to make it eligible to Rust
translation. For this case study, we chose to modify the source \lowstar code rather than
the C that is auto-generated from \lowstar. We did so for two reasons: first, we wanted to preserve
the formal guarantees of \haclstar, rather than perform unverified edits on C code; second,
we wanted to assess how much harder verification would become if we rewrote parts of the library. We
report that the verification impact was negligible, and that our changes
only incurred a few person-days of effort.

We now report on our porting effort for the following algorithms: Chacha20Poly1305, Curve25519,
P256, SHA2, Poly1305, and supporting bignum libraries. These represent different
flavors of code, totalling 14,018 lines of C, and cover almost all usage patterns present in \haclstar.

Chacha20Poly1305 is an authenticated encryption algorithm~\cite{chachapoly}.
Its \haclstar implementation performs no heap allocation and offers a one-shot API. As such,
we were able to translate it to Rust with no modifications, except copying a temporary array to satisfy mutability constraints.

The P256 elliptic curve~\cite{p256-standard} relies on a general-purpose bignum library for core
field operations. For benchmarking, we show a composite P256 benchmark that
relies on the bignum library. P256 is authored in a style where arguments to
core field operations may be disjoint or
equal; in other words,
\li+void fmul(uint32_t *dst, uint32_t *a, uint32_t *b)+,
allows its \li+dst+ and \li+a+ arguments to potentially alias one another; this is typical of
cryptographic code. The main difficulty for P256 was to rewrite such functions
into two variants: one that takes disjoint
arguments, and one that takes in-place arguments, performing a copy in order to
call the disjoint version. Using the equivalent of macros in \lowstar
(helpers marked as \li+inline_for_extraction+), we were able to make this change
transparent from the verification perspective.
We also had to enact several minor, local fixes, mostly to avoid generating code that dereferences
the state (which would generate a move), and instead reference only fields of the state (which Rust
can then auto-borrow).
The Curve25519 elliptic curve~\cite{curve25519} relies on a custom bignum library just for this
particular field.
We encountered the same
flavor of problems as with P256; we followed the same methodology to allow the translation to proceed.

The SHA2 family of hash algorithms~\cite{sha2-standard} offer a ``streaming'' API that relies on long-lived
heap-allocated state, and non-trivial ownership patterns. The original code was written for C, not
Rust, meaning that the code and proofs leveraged aliasing relationships between local variables and
fields of the state. We performed a series of local, targeted rewritings to avoid a few patterns
that were incompatible with Rust. %

The Poly1305 message authentication code (MAC) algorithm~\cite{bernstein2005poly1305}
is also streaming, and furthermore manages ownership of a key throughout
the lifetime of its state. Fortunately, the streaming API of \haclstar is written and
proven once~\cite{ho2023hacl},
meaning that our changes above also benefitted Poly1305; a few additional tweaks were needed for
parts of the code that deal with the key.%

While close to being translatable to Rust, these algorithms still required
some changes. Out of 35,879 lines of (source, \lowstar) code for these algorithms, \li+diff+ reports
1296 insertions, and 318 deletions, meaning 4.5\% of the source code was affected. This diff also
includes fixed proofs.

For performance, we report individual numbers for each algorithm;
performance is largely identical, save for P256 that is significantly faster in
Rust. We speculate that this may be due to link-time optimizations,
which are more aggressive in Rust than in C, and may allow bignum operations to
be composed more efficiently with the P256 code.

\paragraph{Translating Verified Code}
Given that \haclstar' C code already includes formal guarantees of memory safety and functional correctness,
one might wonder, what is the point of translating it to safe Rust instead of using FFI bindings and encapsulating
the library in a Rust crate?
First, having a library entirely written in Rust simplifies the build system, as it directly allows to use
the Rust ecosystem (\li`cargo`) instead of managing multi-language builds. More importantly, avoiding the use
of \emph{unsafe}, needed to write bindings, increases the trust in the library and facilitates adoption of the code:
instead of trusting a (potentially) foreign language such as \fstar or Rocq with the safety of their code,
users can rely on the more familiar \li`rustc` compiler.

\subsubsection{The bzip2 compression library}

All three examples above concern cryptographic libraries, which fundamentally fit within the subset
of code that Scylla was designed to translate (modulo small edits). Going beyond cryptographic
algorithms, we now study the bzip2 compression library~\cite{bzip2}.
This constitutes a relevant example for many
reasons: first, the code is very low-level, meaning that understanding it deeply enough to
rewrite it in Rust manually requires considerable effort.
Second, the code has been hand-tuned and optimized, with a lot of accumulated knowledge -- one
corner case, according to a comment, is triggered by compressing 48.5 million instances of the
character 251. An automated translation will preserve these corner cases. Third, the code is
security-critical, and as such, is a prime candidate for conversion to a safe language.

The library consists of three parts: compression, decompression, and orchestration. Of these
three, Scylla can only translate compression. Decompression is ineligible for translation to safe
Rust, since its main function uses unstructured control-flow in C, with a switch
statement jumping to a label found within a \li+for+-loop. Orchestration interacts deeply
with the operating system through file streams, and would need to be rewritten in a
Rust-idiomatic way. We thus focus on compression only, representing 2969 lines of C code and 34\% of the bzip2
source code (excluding tests and the bzip2recover utility, which we did not attempt to translate).

The bzip2 library is an old piece of code, written in the mid-90s. One particular difficulty was
that its state type contains multiple aliases towards the same piece of data, presumably to allow
for direct loads rather than forcing offset computations -- perhaps this had a measurable
performance impact in the 1990s. In any case, this is exactly the type of pattern that is
incompatible with Rust. This is where our methodology shines: we were able to experiment with
rewrites that eliminated those aliases, while running the (comprehensive) bzip2 test suite to ensure
the C code remained correct. We removed those aliases, relying on
macros to replicate the offset computations at use-site while minimizing changes to the source.
Other familiar issues cropped up: we had to shorten the lifetime of several variables.
In a hot loop, an alias to a mutable array was replaced by offset
computations, potentially generating an indirect load -- again, this may have been important 30
years ago. We also found one unncessary \li+goto+ (unsupported) that could be replaced with a
\li+break+ (supported). Among files related to compression, only 3.5\% of the
source code was impacted. Similarly to FrodoKEM, the translation process unearthed two sources
of undefined behaviors: an alignment issue and a strict aliasing violation. We
reported both issues upstream.

Building bzip2 with our changes thus required a hybrid build, in which compression is in Rust, but
decompression and orchestration remain in C. This required adding a custom FFI to go back and forth
between both languages; furthermore, because the state type contains pointers, a conversion between
C state and Rust state was necessary, the former containing raw pointers and the latter containing
slices. This state conversion is non-trivial, as the state type contains 36 fields. In spite of
this, our benchmark, which compresses 100M random bytes, shows only
a 8\% overhead for our Rust version -- we posit
that this overhead would be largely eliminated if we switched from a hybrid
build to a pure Rust build (thus eliminating state conversions).

\subsubsection{EverParse}

Finally, we turn our attention to another security-critical domain that would benefit from
conversion to safe Rust: parsers and serializers. For our final case study,
we focus on EverParse~\cite{ramananandro2019everparse}, a verified parsers and
serializers library.
We particularly focus on the CBOR-DET parser~\cite{cbor-rfc} generated by EverParse. CBOR-DET is an ongoing IETF draft for a binary format akin to JSON,
and specifically, a deterministic variant of it. It consists
of 4700 lines of C code.

To translate CBOR-DET through Scylla, the key changes were to annotate several tagged union datatypes to reconstruct them as Rust ADTs, and to reconstruct additional datatypes as tuples to pass Rust's mutability analysis.
This required the addition of 19 lines of code. Additionally, we needed to add default values for variables declared and later defined, but where the Rust compiler was not able to
statically determine that the variable was defined for all execution paths. This typically
occurred on patterns of the shape \li{int x; if (...) { x = e } else { exit(...) }},
where one of the execution paths was an error, translated to a Rust panic. While pervasive
(accounting for about 4.1\% of the codebase), these changes were however straightforward.

On average, Scylla-generated code for CBOR is slower than its C counterpart.
However, the performance is highly sensitive to inlining annotations. Because
this code is originally verified, it tends to rely on a myriad of small helper
functions (easier to verify) which get inlined away by the compiler.
Aggressively adding (by hand) inlining annotations makes our Rust version
22\% faster than the C version; however, inlining did not have a noticeable impact on the C code.
This suggests that the differences reflect diverging inlining
strategies between the compilers, rather than a fundamental
limitation with either our approach or \texttt{rustc}.

\subsection{Comparison with c2rust}

To conclude our evaluation, we now compare the output of Scylla with related
tools; in practice, only \texttt{c2rust} can be successfully compiled from its
repository; other tools mentioned in \sref{related} either do not build, cannot process our source code, or
fail to detect any subset of the code that could be translated from unsafe Rust
to Rust.
For this comparison, we rely on the Chacha20 implementation from \haclstar, which we translate with
both Scylla and \texttt{c2rust}. We present in \fref{chacha-compare} the translations of the \li{encrypt_block}
function (shortened for brevity), whose C version is shown below.

\begin{figure*}[!t]
  \footnotesize
  \begin{minipage}{\linewidth}%
    \centering
    \begin{minted}[linenos=false, resetmargins]{rust}
pub fn chacha20_encrypt_block(ctx: &[u32], out: &mut [u8], incr: u32, text: &[u8]) {
  let mut k: [u32; 16] = [0u32; 16usize];
  chacha20_core(&mut k, ctx, incr);
  let mut bl: [u32; 16] = [0u32; 16usize];
  for i in 0u32..16u32 {
    let bj: (&[u8], &[u8]) = text.split_at(i.wrapping_mul(4u32) as usize);
    let u: u32 = load32_le(bj.1);
    let os: (&mut [u32], &mut [u32]) = bl.split_at_mut(0usize);
    os.1[i as usize] = u }
}
    \end{minted}
  \end{minipage}%

  \vspace{1em}
  \hrule
  \vspace{1em}

  \begin{minipage}{\linewidth}%
    \centering
    \begin{minted}[resetmargins,linenos=false]{rust}
unsafe extern "C" fn chacha20_encrypt_block(
    mut ctx: *mut uint32_t, mut out: *mut uint8_t, mut incr: uint32_t, mut text: *mut uint8_t,
) {
  let mut k: [uint32_t; 16] = [0 as libc::c_uint, 0, 0, 0, 0, 0, 0, 0, 0, 0, 0, 0, 0, 0, 0, 0,];
  chacha20_core(k.as_mut_ptr(), ctx, incr);
  let mut bl: [uint32_t; 16] = [0 as libc::c_uint, 0, 0, 0, 0, 0, 0, 0, 0, 0, 0, 0, 0, 0, 0, 0,];
  let mut i: uint32_t = 0 as libc::c_uint;
  while i < 16 as libc::c_uint {
    let mut bj: *mut uint8_t = text.offset(i.wrapping_mul(4 as libc::c_uint) as isize);
    let mut u: uint32_t = load32_le(bj);
    let mut os: *mut uint32_t = bl.as_mut_ptr();
    *os.offset(i as isize) = u;
    i = (i as libc::c_uint).wrapping_add(1 as libc::c_uint) as uint32_t as uint32_t; }
}
    \end{minted}
  \end{minipage}%
  \caption{Translation of encrypt\_block with Scylla (top) and \texttt{c2rust} (bottom)}
  \label{fig:chacha-compare}
\end{figure*}

\begin{minted}[xleftmargin=.5cm, resetmargins]{c}
void chacha20_encrypt_block(uint32_t *ctx, uint8_t *out, uint32_t incr, uint8_t *text) {
  uint32_t k[16U] = { 0U };
  chacha20_core(k, ctx, incr);
  uint32_t bl[16U] = { 0U };
  for (int i = 0; i < 16; i++) {
    uint8_t *bj = text + i * 4U;
    uint32_t u = load32_le(bj);
    uint32_t *os = bl;
    os[i] = u;
}
\end{minted}

While small, this function contains several representative features of C, including
pointer arithmetic to retrieve a subslice (on line 6), structured control-flow,
and array accesses and updates. As previously described, \texttt{c2rust} relies entirely
on unsafe code, which allows the use of C-like pointers in Rust (e.g., \li{*mut uint32_t}).
In contrast, Scylla relies on borrows (e.g., \li{&[u32]}), which come with static memory
safety guarantees. Scylla's mutability inference analysis also allows to mark several
variables and arguments as immutable, while \texttt{c2rust} pervasively uses \li{mut} qualifiers.
However, Scylla's use of safe Rust restricts possible uses of the \li{encrypt_block} function:
in C, this function allowed the use of \emph{in-place encryption}, i.e., passing the same
pointer for arguments \li{out} and \li{text}, as \li{out} is only written after \li{text}
has been read (omitted here). While allowed through the use of unsafe in \texttt{c2rust}'s translation,
the mutability xor aliasing rule of the Rust borrow-checker prevents this for the Scylla
safe version; providing in-place encryption would require a rewrite of the source C program
to expose an API with a single in/out argument. As shown in this example, Rust code generated by
Scylla aims to be readable and maintanable: names from the original C program are preserved
(modulo lexical conventions occasionally differing in C and Rust), and the use of idiomatic
Rust patterns simplifies maintaining and evolving translated codebases.

\section{Alternative Toolchain}

\begin{figure}
  \centering
  \includegraphics[width=.6\textwidth]{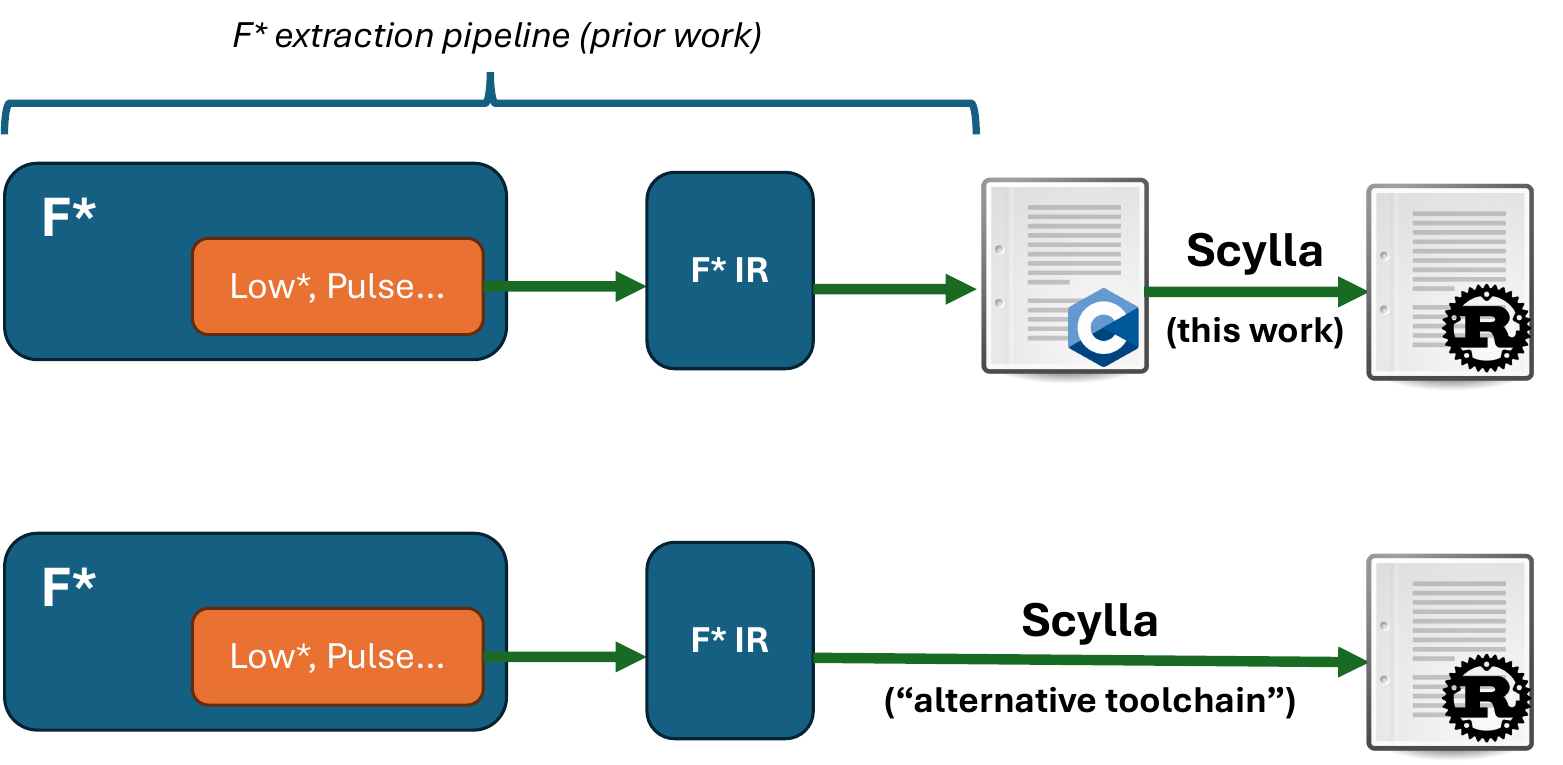}
  \caption{General-purpose toolchain (above), and alternative toolchain (below)}
  \label{fig:alt-toolchain}
\end{figure}

Some of our case studies involve the \fstar language, and specifically, those of
its DSLs that compile (a subset of) \fstar to C. Those are \lowstar~\cite{lowstar} (used by our
\haclstar case study), Steel~\cite{steel2021}, and Pulse~\cite{ebner2025pulsecore} (used by our EverParse case study). In this
paper, we presented a general-purpose toolchain that consumes C code; for those
case studies, this means that we directly consumed C code produced by \fstar,
relying on Scylla to parse it and translate it to Rust.

As an experiment, we developed an alternative toolchain (\fref{alt-toolchain}) that skips concrete
C syntax, and directly consumes \fstar's intermediary representation (IR).
\fstar's IR is isomorphic to \minic, meaning we can emit Rust code directly
without going through the artifact of an actual C file.

This alternative toolchain outputs identical Rust code, but operates more
efficiently. In addition to build simplification and usability improvements, the
alternative toolchain \emph{relieves} the programmer of some of the annotation
burden: because the \fstar IR still has tuples, data types, and in some
cases, a model of slices, this means that the programmer need not
provide annotations for those.

This alternative toolchain was used for a new generation of parsers and
serializers~\cite{ramananandro2025secureparsingserializingseparation}, and for
Bert13~\cite{cryptoeprint:2025/980}, a formally verified implementation of TLS 1.3 which
relies on \haclstar compiled straight from \fstar to Rust.
Naturally, this toolchain has limited applicability, and we advocate
for the usage of Scylla, to be able to translate C code generally, regardless of
whether it was generated from \fstar.

\section{Related Work}
\label{sec:related}

Since this is such a high-value target, many attempts have been made to automatically translate C to
Rust.
Historically, the first widespread tool for C to Rust translation is the aptly-named
C2Rust~\cite{c2rust}, which combines earlier attempts Citrus~\cite{citrus} and
Corrode~\cite{corrode}. The tool translates C99 code to unsafe Rust, leveraging Rust's ability to
manipulate raw, C-like pointers via the use of \li+unsafe+. The tool accepts a large subset of C,
and envisions a manual rewriting of the unsafe Rust into safe Rust.
Several tools were since then built atop C2Rust, trying to automate the rewriting process that
gradually removes \li+unsafe+ in the translated Rust code.

Emre et al.~\cite{emre2021translating} simply borrow-check the output of C2Rust, hijacking
the Rust compiler in order to see if any unsafe raw pointers actually \emph{do} borrow-check. In the
event that they do, the usage of \li+unsafe+ is removed, and the raw C pointers get promoted to
regular, safe Rust borrows.
Their analysis, however, focuses on
\emph{functions whose unsafety comes from usage of raw pointers only}, meaning that in
their case study, \texttt{libxml2}, they rewrite 210 functions out of 3,029,
but with a 97\% success rate.
The rest of the functions remain unsafe.

Zhang et al.~\cite{zhang2023ownership} build a custom set of ownership- and
mutability-based analyses that operate post-C2Rust, again trying to limit the amount of unsafe
pointers and unsafe code. The tool, dubbed Crown, achieves greater unsafe reduction rates
than Emre et al.~\cite{emre2021translating}.

Emre et al.~\cite{emre2023aliasing} later opt for an in-house analysis (rather than
reusing the Rust compiler), which allows for more functions to be successfully translated to safe
Rust. This newer analysis may rewrite programs (rather than simply turn raw pointers
into borrows). Still, only a fraction of the functions
eventually make it to safe Rust.
The authors share the issues with nominal struct types (\S7.2)
that we describe in \sref{mutability}, but do not circumvent it through a translation to tuples
like we do.

Ling et al.~\cite{ling22safer} push the idea further, and use TXL, a
programming language dedicated to source-to-source transformations, to automate the rewriting
of unsafe Rust into safe Rust even more. The authors introduce ``semantics-approximating''
rules, which ultimately allow them to convert a large fraction of unsafe functions into safe ones,
although the preservation of semantics does not seem to be guaranteed. Once again, this is seen as a
stepping stone for engineers to tackle the migration of a codebase, rather than a fully automated
``fire and forget'' translation approach.

To support the translation of C code with unstructured control-flow, C2Rust relies on the
Relooper algorithm to translate gotos to structured control flow~\cite{ramsey22relooper,zakai11emscripten}.
While the resulting Rust code is harder to match with the original C sources, we could implement this
approach in Scylla to extend the C subset we support.

All of the works above aim to translate first into almost fully-unsafe Rust, then either manually or
automatically rewrite the code to pull it out of the unsafe subset, reaching for \emph{partial
safety}.
We advocate for a different approach: iterate on the C code until it successfully translates,
keeping existing integration, unit testing, or proofs to assess the validity and/or correctness of
the rewrites. Then, once the rewrite fits within our supported subset, proceed,
and obtain \emph{fully safe code}.

A separate line of
work takes a more
focused approach and studies specific patterns that occur in C to translate them to
idiomatic Rust -- this is a concern that goes beyond mere safety and correctness. This line of work
still builds atop C2Rust. Problems include: translating the Posix Lock API of C
to Rust's native lock and ownership semantics~\cite{hong2023concrat}; replacing
output parameters with more Rust-native return values and algebraic data types through
an abstract interpretation analysis to infer what constitutes an output
parameter~\cite{hong2023improving,hong2024don}; or reconstructing Rust enumerations
from C tagged unions~\cite{hong2024tag}. Again, these all partially remove
\li+unsafe+; furthermore, these target
a more systems-oriented subset of C, rather than the data-oriented subset we tackle.

Li et al.~\cite{li2025translating} study the process of translating C code to Rust
from a user perspective. They particularly observe that manually translating the code
is error-prone, and that it is important to target idiomatic Rust code. Our approach
aims to follow these guidelines; the readable code outputted can also serve as a basis
for further rewritings.

With the explosion in popularity of LLMs, translating C to Rust with an LLM was inevitable. Pan et al.~\cite{pan2024lost} find that while GPT-4 generates code that is
more idiomatic than C2Rust, only 61\% of it is correct (i.e., compiles and produces
the expected result), compared to 95\% for C2Rust. Hong et al.~\cite{hong2025type}
investigate how to mitigate errors introduced by LLMs in the translation process, while
Yang et al.~\cite{yang2024vert} attempt to validate LLM-translated code using property-based
testing and bounded model checking.
Nitin et al.~\cite{nitin2025c2saferrust} propose to rely on LLMs to make Rust code generated
by C2Rust both safer and more idiomatic. Their approach however only reduces the amount of unsafe
code by up to 28\%.
These works tackle another problem area, where possibly-faulty translations
might be acceptable. We focus instead on critical applications,
where having confidence in the safety of the produced Rust code is paramount.

Focusing on C++ instead, CRAM~\cite{cram} advocates for an approach similar in
spirit to ours: rewrite the source \emph{C++} code to use abstractions that encode Rust's
borrow-checking discipline using advanced C++ templates and type-system features; then, once the
code has been sufficiently refactored, translate it to Rust. Details on CRAM are scarce, as the tool
appears to be closed-source, and no publications are available. The translation is described as ``provably
safe'', which does not conclusively indicate whether the produced code uses unsafe or not.

The tool cpp2rust aims to produce \emph{safe} Rust out of source C++;
it does by producing reference-counted Rust (i.e., code that uses \li+RefCell+).
Subsequent static analyses rewrite the code, wherever possible, to get rid of
the uses of \li+RefCell+.

\section{Conclusion}

We presented a methodology that provides a predictable, deterministic and
auditable translation from a sizeable, data-oriented subset of C to \emph{safe} Rust. Our
methodology relies on an iterative approach: rather than translate to
\emph{unsafe} Rust, we instead improve the C code until it has enough structure
that it becomes eligible for compilation to \emph{safe} Rust. This allows
leveraging existing engineering processes, regression suites, and continuous
integration, to ensure that the rewrites cause no change of behavior. For
security-critical code where even tiny mistakes may incur a very high cost
(e.g., cryptography), we believe this provides a more compelling approach than
existing LLM-based rewrites, or than toolchains that generate unsafe Rust.

The subset of C we support is large enough to translate:
cryptographic algorithms written in different styles; parsers and serializers;
and a large chunk of the well-known \li+bzip2+ compression library. We provide
readable, maintainable code, leveraging safe Rust abstractions such as slices,
enums and matches, rather than producing low-level C code disguised as unsafe
Rust.

Our experimental evaluation shows that the changes required to the original C
code are small, and that the performance of the resulting safe Rust code is
almost always indistinguishable from that of the original C. This provides, we
believe, concrete evidence that our approach scales and provides a pathway for
high-assurance rewrites from C to Rust.

While the various steps of our translation are formalized with pen-and-paper,
and our case studies empirically show that codebases translated by Scylla
pass functional tests, we currently do not have any formal guarantees
about the correctness of the translation. We however hope to mechanically
formalize and prove the correctness of our scheme in future work, by framing
it as a semantic equivalence at the source level.

\begin{acks}
We thank Guillaume Boisseau and Son Ho for several discussions about the Rust semantics
and borrow-checker, Tahina Ramananandro for invaluable help with understanding
EverParse and the CBOR-DET implementation, and the anonymous reviewers for valuable
feedback about earlier versions of this draft.
This work benefited from funding managed by the French National Research Agency under the France 2030
programme with the reference ANR-22-PECY-0006.
\end{acks}

\paragraph{{Data-Availability}}
Our tool is open-source, and publicly available on \href{https://github.com/AeneasVerif/scylla}{Github}.
Additionally, to foster reproducibility, we provide an artifact available on Zenodo which contains
the version of Scylla described in this paper, as well as the experimental evaluation~\cite{scyllaartifact}.

\bibliographystyle{ACM-Reference-Format}
\bibliography{conferences,cited}

\end{document}